\documentclass[aps,amsmath,amssymb,superscriptaddress,prb,10pt,twocolumn,eqrefnum,nofootinbib]{revtex4-2}

\usepackage{graphicx}
\usepackage{dcolumn}
\usepackage{bm}
\usepackage{physics}
\usepackage{color,xcolor}
\usepackage[hyperfootnotes=false]{hyperref}
\usepackage{tikz}
\usepackage{float}
\usepackage[normalem]{ulem}

\renewcommand{\vec}[1]{\mathbf{#1}}

\newcommand{\ch}{\mathrm{charge}}
\newcommand{\spin}{\mathrm{spin}}
\newcommand{\cc}{\mathrm{ch}}
\newcommand{\sn}{\mathrm{sp}}
\newcommand{\upup}{{\uparrow\uparrow}}
\newcommand{\dodo}{{\downarrow\downarrow}}

\begin{document}

\title{Theory of quantum-geometric 
charge and spin Josephson diode effects in strongly spin-polarized hybrid structures with noncoplanar spin textures}

\author{Niklas L. Schulz}
\email{niklas.schulz@uni-greifswald.de}
\affiliation{Institute of Physics, University of Greifswald, Felix-Hausdorff-Strasse 6, 17489 Greifswald, Germany}
\author{Danilo Nikoli\'c}
\email{danilo.nikolic@uni-greifswald.de}
\affiliation{Institute of Physics, University of Greifswald, Felix-Hausdorff-Strasse 6, 17489 Greifswald, Germany}
\author{Matthias Eschrig}
\email{matthias.eschrig@uni-greifswald.de}
\affiliation{Institute of Physics, University of Greifswald, Felix-Hausdorff-Strasse 6, 17489 Greifswald, Germany}

\date{\today}

\begin{abstract}
We present a systematic study of the spin-resolved Josephson diode effect (JDE) in strongly spin-polarized ferromagnets (sFM) coupled to singlet superconductors (SC) via ferromagnetic insulating interfaces (FI). All metallic parts are described in the framework of the quasiclassical Usadel Green's function theory applicable to diffusive systems. The interfaces are characterized by an S-matrix obtained for a model potential with exchange vectors pointing in an arbitrary direction with respect to the magnetization in the sFM. Our theory predicts a large charge Josephson diode effect with an efficiency exceeding $33\%$ and a perfect spin diode effect with $100\%$ efficiency. To achieve these the following conditions are necessary: (i) a noncoplanar profile of the three magnetization vectors in the system and (ii) different densities of states of spin-$\uparrow$ and spin-$\downarrow$ bands in the sFM achieved by a strong spin polarization. The former gives rise to the quantum-geometric phase, $\Delta\varphi$, that enters the theory in a very similar manner as the superconducting phase difference across the junction, $\Delta\chi$. We perform a harmonic analysis of the Josephson current in both variables and find symmetries between Fourier coefficients allowing an interpretation in terms of transfer processes of multiple equal-spin pairs across the two ferromagnetic spin bands. We point out the importance of crossed pair transmission processes. Finally, we study a spin-switching effect of an equal-spin supercurrent by reversing the magnetic flux in a SQUID device incorporating the mentioned junction and propose a way for measuring it.
\end{abstract}

\maketitle


\section{Introduction}\label{sec:Intro}
Superconducting spintronics has the potential for new functionalities of spintronics devices based on equal-spin supercurrents \cite{buzdinProximityEffectsSuperconductorferromagnet2005,bergeretOddTripletSuperconductivity2005,eschrigSymmetriesPairingCorrelations2007,eschrigSpinpolarizedSupercurrentsSpintronics2015,linderSuperconductingSpintronics2015,birgereview2018,linderOddfrequencySuperconductivity2019,yangBoostingSpintronicsSuperconductivity2021,caiSuperconductorFerromagnetHeterostructures2023}. 
Especially promising are devices involving strongly spin-polarized itinerant ferromagnets in a Josephson junction between two spin-singlet superconductors, in which phase-coherence in the ferromagnet is maintained only within each spin band, however not between the two spin bands \cite{eschrigTheoryHalfMetalSuperconductor2003,eschrigTripletSupercurrentsClean2008,greinSpinDependentCooperPair2009,eschrigScatteringProblemNonequilibrium2009,greinInverseProximityEffect2013,eschrigSpinpolarizedSupercurrentsSpintronics2015,houzetQuasiclassicalTheoryDisordered2015,bobkovaGaugeTheoryLongrange2017,ouassouTripletCooperPairs2017,eschrigTheoryAndreevBound2018}.
In superconductors with a conventional order parameter, pair correlations can be classified into singlet and triplet correlations. When penetrating a ferromagnet due to the superconducting proximity effect, the triplet pair correlations become inequivalent and divide into short-range and long-range pairs \cite{bergeretLongRangeProximityEffects2001}. With respect to the ferromagnet's spin quantization axis, equal-spin $\uparrow\uparrow$- and $\downarrow\downarrow$-pair amplitudes are long-ranged in their respective spin band, whereas the third spin-triplet pair amplitude is short-ranged, just like the singlet amplitude, due to the decoherence induced by the exchange splitting of the electronic spin bands. Thus, the building blocks for superconducting spintronics are not the electron spins as in spintronics, but the spins of equal-spin pairs. 

The creation and control of equal-spin supercurrents is based on two fundamental processes taking place near the interface of a superconductor and a ferromagnet: \cite{eschrigSpinpolarizedSupercurrentsSpintronics2011} (i) spin-mixing (or spin-dependent phase shifts) and (ii) triplet rotation. The first process turns spin-singlet pairs into spin-triplet pairs, and the second process turns short-range amplitudes into long-range amplitudes. Whereas the first process merely requires a spin-polarzation of the interface region, the second process requires a noncollinear spin arrangement. Spin-triplet supercurrents based on these mechanisms have been experimentally realized \cite{keizerSpinTripletSupercurrent2006,khaireObservationSpinTripletSuperconductivity2010,anwarLongrangeSupercurrentsHalfmetallic2010,robinsonControlledInjectionSpinTriplet2010,robinsonPRL2010,Wang2010,Gu2010,Sprungmann2010,Glick2018,Caruso2019,Aguilar2020,birge2024}, and at present can be produced routinely.

Whereas long-range triplet supercurrents are a pre-requisite of superconducting spintronics, important new functionalities appear when the spin-texture in the ferromagnet is not only noncollinear, but {\it noncoplanar}. Noncoplanar spin arrangements in strongly spin-polarized superconducting spintronics devices lead to an effective decoupling of the Josephson phases in the two spin bands and an opening of a new channel of control via \mbox{\textit{quantum-geometric phases}} that directly add to the Josephson phases with opposite sign in the two spin bands \cite{greinSpinDependentCooperPair2009,schulzQuantumgeometricSpinCharge2025}. We will show in the following that new phenomena occur in this case, that are entirely governed by quantum-geometric phases that appear due to noncoplanar spin textures in the device. Quantum-geometric phases play an important role in materials with nontrivial spin-structure like altermagnets \cite{Altermagnetism_review1,fengAnomalousHallEffect2022,Altermagnetism_review2}, topological insulators \cite{kaneQuantumSpinHall2005,bernevigQuantumSpinHall2006,hsiehTopologicalDiracInsulator2008,Hasan2010,andoTopologicalInsulatorMaterials2013}, or skyrmionic materials \cite{bogdanovThermodynamicallyStableMagnetic1994,rosslerSpontaneousSkyrmionGround2006,muhlbauerSkyrmionLatticeChiral2009,yuRealspaceObservationTwodimensional2010,Magnetic_skyrmions2017}.
In this article, we discuss the appearance of charge and spin Josephson diode effects based on quantum-geometric phases, as well as a new channel of control for spin-polarized supercurrents, allowing for example for a switching between almost pure spin-up equal-spin supercurrents and spin-down equal-spin supercurrents.

From general considerations, it is well known that if the system is invariant under the time reversal and inversion operations, the normal Josephson effect, characterized by $I(-\Delta\chi)=-I(\Delta\chi)\implies I(\Delta\chi=0)=0$, appears~\cite{golubovCurrentphaseRelationJosephson2004}. However, if the mentioned symmetries are broken this relation can be violated and the so-called "$\varphi_0$-junction" may appear. This effect has been predicted in various setups including junctions with strong spin-orbit coupling~\cite{Buzdin2008}, unconventional superconductors~\cite{GeshkenbeinLarkin1986,Yip1995,Sigrist1998}, or strongly spin-polarized itinerant ferromagnets~\cite{eschrigTripletSupercurrentsClean2008,greinSpinDependentCooperPair2009,bobkovaGaugeTheoryLongrange2017}, to mention a few. Furthermore, the presence of higher harmonics in the Josephson current-phase relation (CPR) leads to nonreciprocal transport. Namely, the critical current in one direction $(+)$ can be different from that flowing in the opposite direction $(-)$ leading to the so-called superconducting or Josephson diode effect (JDE)~\cite{nadeemSuperconductingDiodeEffect2023}. This effect has been intensively studied in numerous experimental \cite{andoObservationSuperconductingDiode2020, baumgartnerSupercurrentRectificationMagnetochiral2022, costaSignReversalJosephson2023, gutfreundDirectObservationSuperconducting2023, houUbiquitousSuperconductingDiode2023, nadeemSuperconductingDiodeEffect2023, strambiniSuperconductingSpintronicTunnel2022, trahmsDiodeEffectJosephson2023} and theoretical works \cite{greinSpinDependentCooperPair2009,costaSignReversalJosephson2023, Margaris_2010,daidoIntrinsicSuperconductingDiode2022, fominovAsymmetricHigherharmonicSQUID2022, haltermanSupercurrentDiodeEffect2022, hePhenomenologicalTheorySuperconductor2022, ilicTheorySupercurrentDiode2022, karabassovHybridHelicalState2022, kopasovGeometryControlledSuperconducting2021, misakiTheoryNonreciprocalJosephson2021, tanakaTheoryGiantDiode2022, yuanSupercurrentDiodeEffect2022, zhangGeneralTheoryJosephson2022, zinklSymmetryConditionsSuperconducting2022,soutoJosephsonDiodeEffect2022,steinerDiodeEffectsCurrentBiased2023,costaMicroscopicStudyJosephson2023,kopasovAdiabaticPhasePumping2023,seoanesoutoTuningJosephsonDiode2024,Meyer2024,ilicSuperconductingDiodeEffect2024,Debnath2024,putilovNonreciprocalElectronTransport2024,tjernshaugenSuperconductingPhaseDiagram2024,patil2024,Cuoco2025} in recent years. The figure of merit in this effect is the so-called diode efficiency defined as 
$ \eta=(I^+-|I^-|)/(I^+ + |I^-|)$
where $I^\pm$ refers to the critical current in the corresponding direction. In the present work, we will focus on two effects of this kind: the charge and the spin Josephson diode effect showing that the former can reach efficiencies exceeding 33\%, whereas the latter can reach an efficiency of up to 100\%. 

To illustrate the role of the noncoplanarity of the magnetization vectors in ferromagnetic trilayers, let us consider the spin rotation from one quantization axis (e.g, $z$-axis) to another one along a direction $\Vec{n}$ parameterized by the polar angle $\alpha$ and the azimuthal angle $\varphi$:
 \begin{equation}
    \begin{pmatrix} \uparrow \\ \downarrow \end{pmatrix}_\Vec{n} = \begin{pmatrix} \cos\frac{\alpha}{2}  &  \sin{\frac{\alpha}{2}e^{i\varphi}} \\ -\sin{\frac{\alpha}{2}e^{-i\varphi}}  & \cos\frac{\alpha}{2} \end{pmatrix} \begin{pmatrix} \uparrow \\ \downarrow \end{pmatrix}_z.
 \end{equation}
 Consequently, the pair amplitudes transform as follows:
 \begin{align}
 \label{eqn:singlet}
     (\uparrow\downarrow-\downarrow\uparrow)_\Vec{n} &=   (\uparrow\downarrow-\downarrow\uparrow)_z,\\
     (\uparrow\downarrow+\downarrow\uparrow)_\Vec{n} &= -\sin\alpha\left[e^{-i\varphi}(\upup)_z-e^{i\varphi}(\dodo)_z\right]\nonumber\\
    \label{eqn:triplet}
   & \quad  +\cos\alpha (\uparrow\downarrow+\downarrow\uparrow)_z.
 \end{align}
 We draw two important conclusions here. First, once mixed-spin triplet correlations [see Eq.~\eqref{eqn:triplet}] are formed along the magnetic moment of the SC/sFM interface (set to be along the $\vec{n}$ direction), it can give rise to equal-spin triplets along the magnetization in the sFM (set to be the $z$-direction). Second, the two equal-spin triplet amplitudes, $(\upup)_z$ and $(\dodo)_z$, acquire a relative phase $\pm(2\varphi+\pi)$ with respect to each other. As we show below, the difference $\Delta \varphi $ of the phase shifts at the two SC/sFM interfaces of the Josephson junction corresponds to the \mbox{\textit{quantum-geometric phase difference}} which in combination with ferromagnetic spin-filtering leads to the appearance of the Josephson diode effect \cite{schulzQuantumgeometricSpinCharge2025}. Moreover, we show that for a given magnetization profile there is a Josephson phase for which the supercurrent is fully spin polarized. Considering a SQUID geometry which incorporates the discussed junction, we propose a device that can switch between nearly fully spin-polarized Josephson currents.
 
 The paper is organized as follows. In Sec.~\ref{sec:QC} we provide general remarks on the quasiclassical Green's function method used for calculating the transport properties. In Sec.~\ref{sec:System} we present a microscopic description of the system under study, consisting of a strongly spin-polarized ferromagnet coupled to two BCS superconductors via ferromagnetic interfaces. In Sec.~\ref{sec:BC} we discuss the boundary conditions which turn out to be crucial for the appearance of the quantum-geometric effects that leads to the Josephson diode effect. In Sec.~\ref{sec:Results} we present the results of our numerical calculations and analyze in detail the Josephson charge and spin diode effects. Finally, in Sec.~\ref{sec:Conclusion} we enclose the discussion by giving the concluding remarks.
\section{Quasiclassical Theory}\label{sec:QC}
For the theoretical description, we make use of the quasiclassical theory of superconductivity of Eilenberger \cite{eilenbergerTransformationGorkovsEquation1968} and Larkin and Ovchinnikov \cite{larkinQuasiclassicalMethodTheory1969} within the limit of diffusive systems, the Usadel theory  \cite{usadelGeneralizedDiffusionEquation1970,belzigQuasiclassicalGreensFunction1999,saulsTheoryDisorderedSuperconductors2022}. This limit is realized in systems where the elastic mean free path $\ell=v_F\tau$ is much smaller than the superconducting coherence length of a ballistic superconductor $\sim\hbar v_F/\Delta_0$ with $v_F$ being the Fermi velocity in the superconductor and $\Delta_0$ the superconducting order parameter at $T=0$. As our main objective is the description of junctions comprising strongly spin-polarized ferromagnets, $J\sim E_F$, the Usadel theory cannot be applied directly, but a modified description is required \cite{greinSpinDependentCooperPair2009,eschrigScatteringProblemNonequilibrium2009,greinInverseProximityEffect2013,bobkovaGaugeTheoryLongrange2017,ouassouTripletCooperPairs2017,eschrigTheoryAndreevBound2018}. 

In this work we focus on equilibrium properties of the system making the consideration of the retarded Green's function $\hat{G}^R(\vec{R},E)$ sufficient. Since we only need this Green's function component, we will omit the $R$ in the exponent in the following, denoting it only as $G(\vec{R},E)$ and referring to it as the Green's function. Note that here $\vec{R}=(x,y,z)^T$ and $E$ denote the spatial center-of-mass coordinate and the energy, respectively. The $\hat{\ldots}$ symbol denotes objects defined in combined particle-hole $\otimes$ spin space spanned by the Gor'kov-Nambu bispinor $\Psi=(\psi_\uparrow,\psi_\downarrow,\psi_\uparrow^\dagger,\psi_\downarrow^\dagger)$. Consequently, the retarded Green's function is a 2$\times$2 matrix in particle-hole space 
\begin{equation}
    \hat{G} = \begin{pmatrix} \mathcal{G} & \mathcal{F} \\ -\tilde{\mathcal{F}} & -\tilde{\mathcal{G}} \end{pmatrix}, \label{eq:ret_GF}
\end{equation}
where each entry itself is a $2\times2$ matrix in spin space. Note that $\tilde{\ldots}$ denotes a particle-hole symmetry relation $\tilde{\mathcal{Q}}(\vec{R},E) = \mathcal{Q}(\vec{R},-E^*)^*$ \cite{sereneQuasiclassicalApproachSuperfluid1983,eschrigDistributionFunctionsNonequilibrium2000}. The matrix structure in spin space can be further decomposed as \cite{sereneQuasiclassicalApproachSuperfluid1983}
\begin{equation}
    \hat{G} = \begin{pmatrix} \mathcal{G}_0 \mathit{1} + \bm{\mathcal{G}} \cdot \bm{\sigma} & (\mathcal{F}_0\mathit{1} + \bm{\mathcal{F}} \cdot \bm{\sigma}) i \sigma_y  \\  -i \sigma_y  (\tilde{\mathcal{F}}_0\mathit{1} + \tilde{\bm{\mathcal{F}}} \cdot \bm{\sigma}) & -(\tilde{\mathcal{G}}_0\mathit{1} + \tilde{\bm{\mathcal{G}}} \cdot \bm{\sigma}^*) \end{pmatrix} \label{eq:ret_GF_spin_decomp}
\end{equation}
where $\bm{\sigma} = (\sigma_x,\sigma_y,\sigma_z)^T$ is the Pauli vector in spin space, $\mathit{1}=\mbox{diag} (1,1)$ is the identity matrix in spin space, $\mathcal{F}_0$ and 
$\bm{\mathcal{F}} = (\mathcal{F}_x,\mathcal{F}_y,\mathcal{F}_z)^T$ are the singlet and triplet pair amplitudes, respectively, and $\mathcal{G}_0$ and $\bm{\mathcal{G}} = (\mathcal{G}_x,\mathcal{G}_y,\mathcal{G}_z)^T$ are the spin-scalar and spin-vector contributions to $\mathcal{G}$. The Green's function is determined by solving the Usadel equation with the normalization condition \cite{usadelGeneralizedDiffusionEquation1970,belzigQuasiclassicalGreensFunction1999,saulsTheoryDisorderedSuperconductors2022},
\begin{equation}
    [E \hat{\tau}_3 - \hat{\Delta},\hat{G}] - i\hbar \sum_{k,l}D_{kl} \nabla_k (\hat{G}\nabla_l\hat{G}) = \hat{0}, \, \hat{G}^2 = \hat{1}, \label{eq:Usadel}
\end{equation}
where $\hat{\tau}_3 = \text{diag}(\mathit{1},-\mathit{1})$ is the third Pauli matrix in particle-hole $\otimes$ spin space, $k,l\in \{x,y,z\}$, and the gap matrix $\hat{\Delta}$ for spin-singlet BCS pairing is given by
\begin{equation}
    \hat{\Delta} = \begin{pmatrix} 0 & \Delta \\ \Delta^* & 0 \end{pmatrix}, \label{eq:self_energy}
\end{equation}
where the entries have a simple spin structure, $\Delta\propto i\sigma_y$. To obtain the full solution, the spatially varying superconducting gap should be calculated self-consistently
\begin{equation}\label{eq:gap_sc}
    \Delta(\Vec{R}) = \lim_{c\to \infty} \frac{-\int_{-c}^c \frac{dE}{2} \mathcal{F}_0(\vec{R},E) \tanh{\frac{E}{2k_BT}} }{\ln{\frac{T}{T_c}} + \int_{-c}^c \frac{dE}{2E} \tanh{\frac{E}{2k_BT}}} i\sigma_y,
\end{equation}
where $T_c$ is the superconducting critical temperature. Finally, $D_{kl}$ is the diffusion coefficient which is, in general, a tensor. However, for an isotropic Fermi surface and electron-impurity scattering as considered in this paper, it takes a simple form $D_{kl}=D\delta_{kl}$ where $D=v_F^2\tau/3$, with $\tau $ the impurity scattering time. 

The Green's function formalism furthermore allows to express the expectation values of observables in a compact form. In this work, we mainly focus on the charge and spin currents in quasi-one-dimensional wires, which in terms of the Usadel Green's functions for spin-degenerate diffusion constants ($D$) and densities of states at the Fermi level ($N_F$) are given by, respectively, 
\begin{align}
    \label{eq:I_ch}
    &I_\ch = eN_FD\, \mathrm{Re}\!\int\limits_{-\infty}^\infty \frac{dE}{4} \mathrm{Tr}\left[\hat{\tau}_3\hat{G}\partial_{z} \hat{G}\right] \tanh{\frac{E}{2k_BT}},\\
    \label{eq:I_sp}
    &\bm{I}_\spin = \frac{\hbar}{2}N_FD\, \mathrm{Re}\!\int\limits_{-\infty}^\infty \frac{dE}{4} \mathrm{Tr}\left[\hat{\tau}_3 \hat{\bm{\sigma}} \hat{G}\partial_{z} \hat{G}\right] \tanh{\frac{E}{2k_BT}}.
\end{align}
Here, $e=-|e|$ is the electron charge, $z$ is the spatial coordinate in positive current direction, $\hat{\bm{\sigma}} = (\hat\sigma_x,\hat\sigma_y,\hat\sigma_z)^T$ with $\hat{\sigma}_i = \text{diag}(\sigma_i,\sigma_i)$ with $i= x,y,z$, and the trace is performed over the entire $4\times4$ particle-hole $\otimes$ spin space.
\subsection{Coherence functions}
In the following, we consider spin-dependent phenomena arising in the vicinity of superconductor-ferromagnet interfaces and thereby we must maintain the full spin structure of the problem. To do so, we utilize the so-called Riccati parameterization \cite{schopohlQuasiparticleSpectrumVortex1995,schopohlTransformationEilenbergerEquations1998,nagatoTheoryAndreevReflection1993,higashitaniMeissnerEffectNormalSuperconducting1995,eschrigElectromagneticResponseVortex1999,eschrigDistributionFunctionsNonequilibrium2000,eschrigSingletTripletMixingSuperconductor2004}, where the Green's function is expressed in terms of coherence (Riccati) functions $\gamma$ and $\tilde{\gamma}$ as
\begin{equation}
    \hat{G} = \hat{N} \begin{pmatrix} (\mathit{1}+\gamma\tilde{\gamma}) & 2\gamma \\ -2\tilde{\gamma} & -(\mathit{1}+\tilde{\gamma}\gamma) \end{pmatrix}, \label{eq:Riccati_param_GF}
\end{equation}
with $\hat{N}$ being the normalization matrix given by \cite{eschrigDistributionFunctionsNonequilibrium2000}
\begin{equation}
    \hat{N} = \begin{pmatrix} (\mathit{1}-\gamma\tilde{\gamma})^{-1} & 0 \\ 0 & (\mathit{1}-\tilde{\gamma}\gamma)^{-1} \end{pmatrix} = \begin{pmatrix} N & 0 \\ 0 & \tilde{N} \end{pmatrix}.
\end{equation}
In general, the coherence functions $\gamma$ and $\tilde{\gamma}$ are $2\times2$ matrices in spin space obeying the following equations of motion [see Eq.~\eqref{eq:Usadel}]~\cite{eschrigSingletTripletMixingSuperconductor2004,konstandinSuperconductingProximityEffect2005,cuevasProximityEffectMultiple2006}:
\begin{align}
    \pdv[2]{\gamma}{z} + \pdv{\gamma}{z} \tilde{\mathcal{F}} \pdv{\gamma}{z} &= \frac{i}{\hbar D} \left[ \gamma\Delta^*\gamma - 2 E \gamma - \Delta\right] \label{eq:usadel_eq1}, \\
    \pdv[2]{\tilde{\gamma}}{z} + \pdv{\tilde{\gamma}}{z}\mathcal{F} \pdv{\tilde{\gamma}}{z} &= \frac{-i}{\hbar D} \left[ \tilde{\gamma}\Delta\tilde{\gamma} + 2 E \tilde{\gamma} -\Delta^*\right] \label{eq:usadel_eq2},
\end{align}
where $\mathcal{F}$ and $\tilde{\mathcal{F}}$ can be read off from Eqs.~\eqref{eq:ret_GF} and (\ref{eq:Riccati_param_GF}). As the equations above constitute two coupled second-order differential equations, the full solution requires appropriate boundary conditions. For that purpose, we utilize the general boundary conditions based on the microscopic S-matrix approach suited for diffusive systems that have been developed in Ref.~\cite{eschrigGeneralBoundaryConditions2015}. However, before we discuss them, let us provide a brief overview of the Usadel theory for the case of strong spin splitting, which turns out to be crucial for the geometric effects presented in this work.
\subsection{Strongly spin-polarized materials}
\begin{figure}[t!]
    \centering
    \includegraphics[width=\linewidth]{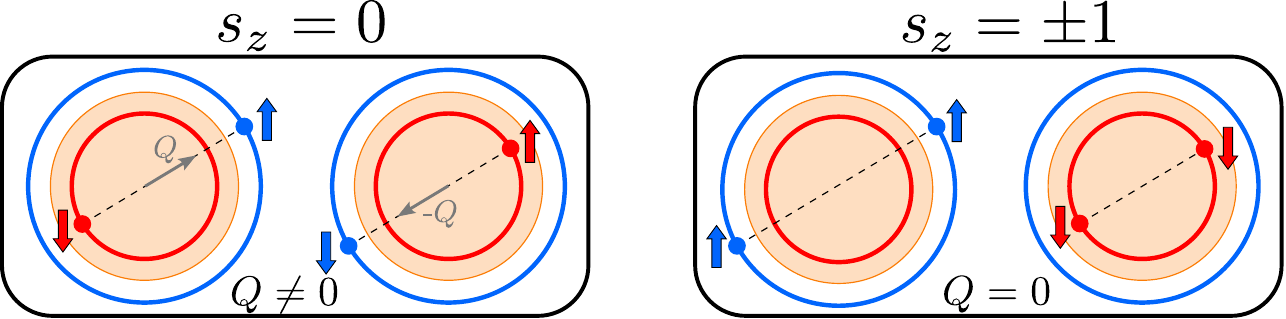}
    \caption{Schematic representation of the two different pairing mechanism for spin-triplet correlations. The left panel shows pairing which leads to mixed-spin triplet correlations (with spin projection $s_z = 0$) with finite center-of-mass momentum $Q \neq 0$. The right panel shows the equal-spin triplet correlations (with spin projection $s_z = \pm 1$), which are generated in a strongly spin-polarized ferromagnet and do not acquire a finite center-of-mass momentum, i.e. $Q=0$. Only the latter ones contribute to the transport across the metallic ferromagnet in our model.}
    \label{fig:pairing}
\end{figure}

Ferromagnetism can be included in the quasiclassical theory by different means. For the limit of weak spin polarization, i.e., $J \lesssim 0.1 E_F$, the exchange field can be directly incorporated in the Usadel equation~[see Eq.~\eqref{eq:Usadel}] as a self-energy that effectively leads to a spin-dependent energy shift $E \rightarrow E - \bm{J}\cdot\bm{\sigma}$ \cite{AORT85,tokuyasuProximityEffectFerromagnetic1988,RosarioFazio99}. However, for strongly spin-polarized materials that exceed this limit an alternative approach is needed. Such an approach was proposed in Ref.~\cite{greinSpinDependentCooperPair2009}, and is based on the consideration of the strong spin polarization prior to the quasiclassical approximation. In this case one defines two separate Green's function for the two spin bands maintaining the phase coherence within each of them but not between them. Only equal-spin ($\upup$ and $\dodo$) correlations with respect to a quantization axis parallel to $\bm{J}$ are kept whereas the mixed ones are neglected. The validity of such an approximation can be justified by considering the relevant length scales in mesoscopic superconductivity. Namely, as it is known, the mixed-spin correlations with total spin projection $s_z=0$ exponentially decay on length scales set by the ferromagnetic exchange length $\sim\sqrt{\hbar D / J}$ which for strong spin polarization, $J\sim E_F$, approaches the atomic length scale characterized by the Fermi wavelength $\lambda_F$. On the other hand, the relevant length scale for equal-spin triplet correlations with the total spin projection $s_z=\pm 1$ is the thermal coherence length, $\sim\sqrt{\hbar D/ k_B T_c}$, which is substantially larger than the exchange length mentioned above. A schematic representation of the above discussion is shown in Fig.~\ref{fig:pairing}, where the left panel displays the \textit{interband} pairing that leads to short-range triplets (those with $s_z=0$) acquiring a finite center-of-mass momentum due to the spin-mixing effects. The right panel, however, shows the \textit{intraband} pairing that results in the equal-spin triplets (those with $s_z=\pm 1$), where both electrons belong to the same spin band, consequently, acquiring zero center-of-mass momentum. In our model, the charge and spin supercurrents within the ferromagnet are entirely carried by the latter ones.

Taking the above assumptions into account and neglecting the mixed-spin amplitudes in the Green's function [see Eq.~\eqref{eq:ret_GF}], we reduce our problem to two decoupled scalar problems. Namely, each spin band is treated separately and described by a $2\times2$ quasiclassical Green's function defined as
\begin{equation}
    \breve{G}_{\eta\eta} = \begin{pmatrix} \mathcal{G}_{\eta\eta} & \mathcal{F}_{\eta\eta} \\ -\tilde{\mathcal{F}}_{\eta\eta} & -\tilde{\mathcal{G}}_{\eta\eta} \end{pmatrix},
\end{equation}
where $\eta = \uparrow,\downarrow$ and $\breve{\ldots}$ denotes a matrix structure in particle-hole space only. Consequently, the propagators for different spin bands of a ferromagnet obey a diffusive motion described by two decoupled Usadel equations 
\begin{equation*}
    [E \breve{\tau}_3,\breve{G}_{\eta\eta}] - i\hbar D_{\eta} \pdv{z} (\breve{G}_{\eta\eta}\pdv{z}\breve{G}_{\eta\eta}) = \breve{0}, \ \breve{G}_{\eta\eta}^2 = \breve{1}.
\end{equation*}
Note that the gap matrix vanishes since there is no BCS pairing in the ferromagnet and, in contrast to the previous considerations, the Fermi velocities and the densities of states at the Fermi level are spin-dependent, as is the diffusion constant $D_\eta $.

Finally, the transport equation for the coherence functions [see Eqs.~\eqref{eq:usadel_eq1} and~\eqref{eq:usadel_eq2}] follows as:
\begin{equation}
    \pdv[2]{\gamma_{\eta\eta}}{z} + \pdv{\gamma_{\eta\eta}}{z} \tilde{\mathcal{F}}_{\eta\eta} \pdv{\gamma_{\eta\eta}}{z} = \frac{-i}{\hbar D_\eta} 2 E  \gamma_{\eta\eta}, \label{eq:usadel_sFM}
\end{equation}
and the equation for $\tilde{\gamma}_{\eta\eta}$ is obtained by applying the $\tilde{\ldots}$-operation to the above equation. 

Since the ferromagnet (sFM) is characterized by two independent Green's function it is convenient to decompose the currents [see Eqs.~\eqref{eq:I_ch} and \eqref{eq:I_sp}] into their contributions from each spin-band,
\begin{equation*}
    I_\ch = 2e(I_{\uparrow\uparrow} + I_{\downarrow\downarrow}), \quad I_\spin = \hbar(I_{\uparrow\uparrow} - I_{\downarrow\downarrow}),
\end{equation*}
and define $I_\cc = I_\upup + I_\dodo$ and $I_\sn = I_\upup-I_\dodo$. Here, the exchange field of the sFM sets the global quantization axis and $I_{\eta\eta}$ is given by
\begin{equation}
    I_{\eta\eta} = N_{F\eta} D_\eta \mathrm{Re}\int\limits_{-\infty}^\infty \frac{dE}{8} \mathrm{Tr}\left\{\breve{\tau}_3 \breve{G}_{\eta\eta}\partial_z \breve{G}_{\eta\eta}\right\} \tanh{\frac{E}{2k_BT}},
\end{equation}
with Tr denoting the trace over particle-hole space only. 
Finally, in terms of the coherence functions, the trace in the expression above can be rewritten as \cite{Gelhausen16}
\begin{equation}
    2 \left[\big\{
    N \left(\partial_z\gamma\right) \tilde N ,\tilde{\gamma}\big\} -\big\{
    \tilde N \left(\partial_z\tilde{\gamma}\right)  N ,\gamma\big\}\right]_{\eta\eta},
\end{equation}
where the subscript $\eta\eta$ indicates that all symbols refer  to spin-band $\eta$, and $\{.,.\}$ is the anticommutator.
\section{System under study}\label{sec:System}
\begin{figure}[t!]
    \centering
    \includegraphics[width =\linewidth]{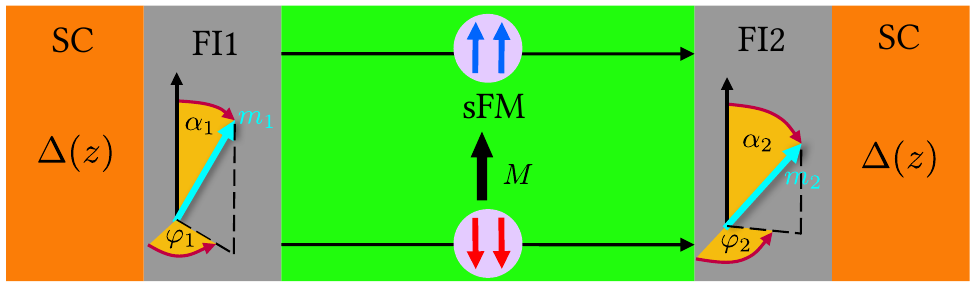}
    \caption{Sketch of the system under study where two superconductors (SC; orange) are brought into contact with a strongly spin-polarized ferromagnet (sFM; green) via two ferromagnetic insulating layers (FI1/FI2; grey). At the outer interfaces the system is connected to superconducting reservoirs characterized by the respective pair potential $\Delta_{1/2} = \abs{\Delta_{1/2}} e^{i\chi_{1/2}}$, where 1/2 corresponds to the left/right interface.$^1$}
    \label{fig:Trilayer_sFM_sketch}
\end{figure}
We study a  superconducting heterostructure consisting of a strongly spin-polarized ferromagnet (sFM) sandwiched between two superconducting leads (SC), separated by ferromagnetic insulating layers ($\mathrm{FI}_{1/2}$) as sketched in Fig.~\ref{fig:Trilayer_sFM_sketch}. At the outer interfaces the superconducting leads are linked to two bulk superconductors via highly transparent interfaces, defining the phase bias across the junction. A microscopic description of each magnetically active region is presented in the following.

\footnotetext[1]{The vectors $\bm{M}$, $\bm{m}_1$, and $\bm{m}_2$ are here defined in units of the quasiparticle magnetic moment (which can be negative for spin-up as, e.g. in Fe, Co, Ni), i.e. they point in the directions of the respective exchange fields.}

\subsection{Strongly spin-polarized ferromagnets}
\begin{figure}[t]
    \centering
    \includegraphics[scale=0.9]{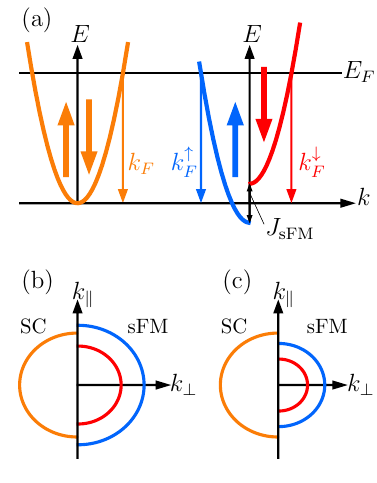}
    \caption{(a): Schematic representation of the parabolic band model. The SC (left; orange) has a spin-degenerate Fermi surface, whereas the sFM (right) has a separate Fermi surface for each spin band (blue $\uparrow$, red $\downarrow$). Two possible Fermi surface geometries are (b) $k_F^\downarrow < k_F < k_F^\uparrow$ and (c) $k_F^\downarrow < k_F^\uparrow < k_F$.}
    \label{fig:sketch_parabolas}
\end{figure}

All ferromagnetic parts are modelled by parabolic bands with spin splitting, i.e., the ferromagnetic order in each subsystem is characterized by a respective exchange field $\bm{J}$ and a bias potential $V$. Therefore, the single-particle dispersion relation in the ferromagnet reads
\begin{align}
    \xi_{\eta }(\Vec{p}) &= \frac{\Vec{p}^2}{2m} - E_F + V_\mathrm{sFM} \mp J_\mathrm{sFM}/2, \label{eq:dispersion_sFM}
\end{align}
where $\eta = \uparrow,\downarrow$, and the upper sign corresponds to $\eta=\uparrow $. 
Since we consider strongly spin-polarized ferromagnets, an important aspect at the SF interfaces is the Fermi surface mismatch created by the strong spin splitting. For definiteness, we consider spherical Fermi surfaces but in general this model can be straight-forwardly extended to non-spherical ones. Sketches of the dispersion relations and the Fermi surfaces of an SC/sFM interface are shown in Fig.~\ref{fig:sketch_parabolas}.
This model allows to express the Fermi velocity of each spin band in the sFM in terms of the Fermi velocity of the superconductor
\begin{equation}
    v_{F,\eta} = v_F \cdot \sqrt{1 - (V_\mathrm{sFM}\mp J_\mathrm{sFM}/2)/E_F},
\end{equation}
which defines  $D_\eta=v_{F,\eta}^2\tau/3$ in Eq.~\eqref{eq:usadel_sFM}.

\subsection{Spin-polarized interfaces}
\begin{figure}[t]
    \centering
    \includegraphics[width=0.7\linewidth]{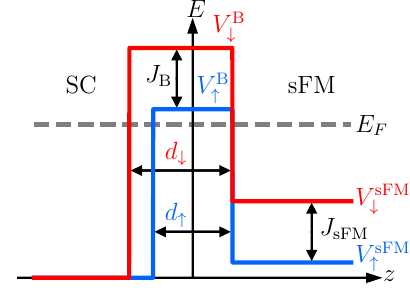}
    \caption{A schematic representation of the potentials characterizing the SC/sFM interface in our model. In the superconductor (SC) it assumed that no bias potentials and exchange fields are present. In the ferromagnetic insulator both spin bands are shifted above the Fermi energy and are therefore insulating whereas in the strongly spin-polarized ferromagnet the bands are splitted but metallic. Since the splitting in both magnetic materials is of the order of $E_F$ the Fermi surfaces are split, see Fig.~\ref{fig:sketch_parabolas}. The potentials in the two regions for each spin direction are given by $V^\mathrm{B/sFM}_{\uparrow(\downarrow)} = V^\mathrm{B/sFM} \mp J^\mathrm{B/sFM}/2$, where it is ensured that $V^\mathrm{B}_{\uparrow(\downarrow)} > E_F$ and $V^\mathrm{sFM}_{\uparrow(\downarrow)} < E_F$.}
    \label{fig:sketch_barrier}
\end{figure}

We now discuss the spin-active insulating layers between the superconductors and the central ferromagnetic region. It is important to emphasize that the ferromagnetic insulating layers are of thicknesses and energies comparable to the Fermi wave length $\lambda_F$ and the Fermi energy $E_F$, respectively, and therefore they enter quasiclassical theory as boundary conditions~\cite{tokuyasuProximityEffectFerromagnetic1988,fogelstromJosephsonCurrentsSpinactive2000,cuevasQuasiclassicalDescriptionTransport2001,kopuTransfermatrixDescriptionHeterostructures2004,zhaoNonequilibriumSuperconductivitySpinactive2004,eschrigScatteringProblemNonequilibrium2009}, which for diffusive systems and arbitrary spin splittings were derived in Ref.~\cite{eschrigGeneralBoundaryConditions2015}. Thus, the interfaces are characterized by the corresponding normal-state scattering matrix connecting the Bloch waves on the two sides of the interface~\cite{lambertGeneralizedLandauerFormulae1991,beenakkerQuantumTransportSemiconductorsuperconductor1992,takaneConductanceNormalSuperconductorContacts1992}. The S-matrix approach is fully quantum and the microscopic model of the interface region is similar to that for the ferromagnetic layer. However, the key distinction is that the interfaces are insulating, i.e., $E_F < V_\mathrm{B} \mp J_\mathrm{B}/2$. The FI layers are modeled as box-shaped potentials schematically shown in Fig.~\ref{fig:sketch_barrier}. Consequently, the transparency of each spin-channel is fully determined by its spin-dependent thickness $d_{\eta}$ and bias potential $V^\mathrm{B}_{\eta}$. In addition, the assumption of spin-dependent widths is a first-order approximation to the case of smooth barrier potentials, see Ref. \cite{greinTheorySuperconductorferromagnetPointcontact2010}, allowing for a higher spin-mixing angle on the superconducting side of the interface. Due to the spin-splitting, the FI layer possess a nonvanishing magnetic moment pointing, in general, in an arbitrary direction with respect to the global quantization axis set by the magnetization vector in the sFM.\footnotemark[1] Accordingly, we parameterize the magnetic moment direction by the spherical angles $\alpha$ and $\varphi$ as shown in Fig. \ref{fig:Trilayer_sFM_sketch}. Consequently, the ferromagnetic insulating layers are fully described by the set of following parameters: $V, J, d_{\uparrow,\downarrow}, \alpha$, and $\varphi$. 
Finally, the S-matrix of an SC/sFM interface discussed above has a general 4$\times$4 matrix structure
\begin{equation}
    \arraycolsep=1pt\def\arraystretch{1.5}
   \bm{S} = \left( \begin{array}{c|c|c} R_\mathrm{SC} & T_\uparrow^{\mathrm{SC} \leftarrow \mathrm{sFM}} & T_\downarrow^{\mathrm{SC} \leftarrow \mathrm{sFM}} \\
    \hline T_\uparrow^{\mathrm{sFM} \leftarrow \mathrm{SC}} & r_\uparrow & r_{\uparrow\downarrow}\\ \hline T_\downarrow^{\mathrm{sFM}\leftarrow \mathrm{SC}} & r_{\downarrow\uparrow} & r_\downarrow \end{array} \right),
\end{equation}
where $R_\mathrm{SC}$ is a 2$\times$2 block describing reflection processes (normal and spin-flip) on the superconducting side. Transmission events are described by the row- ($T_\uparrow^{\mathrm{sFM} \leftarrow \mathrm{SC}}$,$T_\downarrow^{\mathrm{sFM} \leftarrow \mathrm{SC}}$) or column-vectors ($T_\uparrow^{\mathrm{SC}\leftarrow \mathrm{sFM}}$,$T_\downarrow^{\mathrm{SC}\leftarrow \mathrm{sFM}}$), which characterize transmission of Bloch waves from region $A$ to $B$ ($B \leftarrow A$). Finally, the reflection processes on the sFM side are characterized by $r_\eta$ and $r_{\eta\overline{\eta}}$.

\section{Boundary conditions}\label{sec:BC}
For solving a specific problem the Usadel equation (\ref{eq:Usadel}) needs to be supplemented with boundary conditions appropriate for the system at hand. As previously mentioned, the case of ferromagnetic insulating barrier layers can be implemented by considering the general boundary conditions derived in Ref.~\cite{eschrigGeneralBoundaryConditions2015} for which we utilise the scattering matrix calculated in the last section. In what follows, we outline the three-set procedure described in detail in Ref. \cite{eschrigGeneralBoundaryConditions2015} but use a modified notation which yields a simplification of the equations therein.

Since we assume the conservation of momentum parallel to the interface [see Fig.~\ref{fig:sketch_parabolas}] we can label the scattering channels by their respective parallel momentum $p_\parallel$. Consequently, all quantities in the following are functions of $p_\parallel$, but we omit the explicit dependence when it is not explicitly needed. Following Ref.~\cite{eschrigGeneralBoundaryConditions2015}, we first decompose $\bm{S}$ into a polar decomposition
\begin{equation}
    \bm{S} = \begin{pmatrix} \sqrt{1 - C C^\dagger} & C \\ C^\dagger & -\sqrt{1-C^\dagger C} \end{pmatrix} \begin{pmatrix} S & 0 \\ 0 & \underline{S} \end{pmatrix},
\end{equation}
where the entries on the right hand side are matrices in spin space. The block diagonal matrix $\text{diag}(S,\underline{S})$, represents an auxiliary scattering matrix for an impenetrable interface and $C$ is a generalised transmission matrix. In the following, we derive a more compact version of the boundary conditions in Ref.~\cite{eschrigGeneralBoundaryConditions2015} by utilizing the principal matrix square root of $S$, which we denote by 
$\mathcal{S} = \sqrt{S}$, and similarly $\underline{\mathcal{S}} =\sqrt{\underline{S}}$.

We parameterize $C$ by a hopping matrix $\tau $ through the barrier as follows:
\begin{equation}
    C = \mathcal{S} \, (\mathit{1}+\tau \tau^\dagger)^{-1} 2\tau \, \underline{\mathcal{S}}^\dagger.
\end{equation}
Hopping in the opposite direction is given by $\underline{\tau}=\tau^\dagger $.
The particle-hole structure of the scattering and hopping matrices are, respectively,
\begin{align}
    \hat{\mathcal{S}}(p_\parallel) &= \begin{pmatrix} \mathcal{S}(p_\parallel) & 0 \\ 0 & \mathcal{S}^T(-p_\parallel) \end{pmatrix}, \\
    \hat{\tau}(p_\parallel) &= 
    \begin{pmatrix} \tau(p_\parallel) & 0 \\ 0 & \underline{\tau}^T(-p_\parallel) \end{pmatrix}. 
\end{align}

The derivation in Ref.~\cite{eschrigGeneralBoundaryConditions2015} assumes separation of the space around the scattering region into a diffusive zone, an isotropization zone, and a ballistic region directly around the scattering region. 
The procedure proceeds by implementing an idea based on Refs.~\cite{cuevasQuasiclassicalDescriptionTransport2001} and \cite{thunebergQuasiclassicalTheoryIons1981}, where it was shown that a scattering problem at a perturbation field can be solved by using an auxiliary Green's function $\hat{g}_0$. This Green's function is the solution of the problem in the absence of the perturbation field but the self-energies are the same as before. In our case such a situation is realized by considering an impenetrable interface allowing us to calculate the transmission matrix $\hat{t}$. To do so, we first need to account for $\hat{g}_0$ which in the considered case can be directly calculated from the diffusive propagator $\hat{G}$. We first introduce two new diffusive propagators
\begin{equation*}
    \hat{G}_1 = \hat{\mathcal{S}} \hat{G} \hat{\mathcal{S}}^\dagger \quad \text{and} \quad \hat{G}_2 = \hat{\mathcal{S}}^\dagger \hat{G} \hat{\mathcal{S}}.
\end{equation*}
This allows to express $\hat{g}_0$ as follows~\cite{eschrigGeneralBoundaryConditions2015}:
\begin{equation}
    \hat{g}_0 = 2 (\hat{1} + \hat{G}_1 \hat{G}_2)^{-1} (\hat{1} + \hat{G}_1) - \hat{1}.
\end{equation}
Note that even though $\hat{G}$ does not depend on the momentum of the incident particle, $\hat{G}_{1,2}$ as well as the ballistic auxiliary propagator $\hat{g}_0$ are functions of $p_\parallel$, which is encoded in $\hat{\mathcal{S}}$. 

The next step is to calculate the $t$-matrix which allows to connect the ballistic propagators on either side of the interface. The $t$-matrix follows from the solution of a Dyson-like equation for $\hat{g}_0$ and reads
\begin{equation}
    \hat{t} = (\hat{1} + \hat{g}_1\hat{g}_0)^{-1} \hat{g}_1,
\end{equation}
where $\hat{g}_1 = \hat{\tau} \underline{\hat{g}}_0 \hat{\tau}^\dagger = \underline{\hat{\tau}}^\dagger \underline{\hat{g}}_0 \underline{\hat{\tau}}$. Finally, $\hat{t}$ is used for calculating the incoming propagator, $\hat{g}_i$, and outgoing propagator, $\hat{g}_o$, on one side of the scattering region
\begin{align}
    \hat{g}_i &= \hat{\mathcal{S}}^\dagger \left[\hat{g}_0 - (\hat{g}_0 - \hat{1}) \hat{t} (\hat{g}_0 + \hat{1})\right] \hat{\mathcal{S}}, \\
    \hat{g}_o &= \hat{\mathcal{S}} \left[\hat{g}_0 - (\hat{g}_0 + \hat{1}) \hat{t} (\hat{g}_0 - \hat{1})\right] \hat{\mathcal{S}}^\dagger.
\end{align}
The difference between these is referred to as the matrix current $\hat{\mathcal{I}}$, which characterizes the transmission and reflection of correlations through/at the barrier, $\hat{\mathcal{I}} = \hat{g}_o - \hat{g}_i$. As introduced in 
Ref.~\cite{nazarovNovelCircuitTheory1999}, the boundary conditions for the diffusive propagators are 
\begin{equation}
    \mathcal{G}_q \sum_{n=0}^{\mathcal{N}_{\text{max}}} \hat{\mathcal{I}}_{nn} = -\frac{\sigma \mathcal{A}}{\xi} \hat{G} \pdv{z} \hat{G},
\end{equation}
were $n$ is the channel index, $\mathcal{G}_q  = e^2 / h$ the quantum of conductance, $\sigma = e^2 N_F D$ is the normal state conductivity per spin, $\mathcal{A}$ the surface area of the contact, $\xi$ the corresponding coherence length of the system, and $z$ the coordinate parallel to the interface normal. In the limit of strong polarization $\sigma$ and $\xi$ become spin-dependent quantities which can be accounted for by redefining them as matrices acting only in spin space. The transition from discrete channel index to the conserved parallel momentum $k_\parallel$ can be done by \cite{eschrigGeneralBoundaryConditions2015}
\begin{equation}
    \frac{1}{\mathcal{N}_{\text{max}}} \sum_{n=0}^{\mathcal{N}_{\text{max}}} \ldots = \frac{1}{A_{F,z}} \int_{A_{F,z}} d^2 k_\parallel \ldots.
\end{equation}
This relation allows to rewrite the boundary conditions as 
\begin{equation}
    \hat{\overline{\mathcal{I}}}\equiv \langle \hat{\mathcal{I}}(k_\parallel) \rangle_{A_{F,z}} = - \frac{\mathcal{G}_N}{\mathcal{G}_B} \hat{G} \pdv{z} \hat{G} = -r \hat{G} \pdv{z} \hat{G}, \label{eq:final_boundary_cond_GF}
\end{equation}
where $\langle \ldots \rangle_{A_{F,z}} = A_{F,z}^{-1} \int_{A_{F,z}} d^2 k_\parallel \ldots$, $A_{F,z}$ is the projection of the Fermi surface onto the plane perpendicular to the barrier, $\mathcal{G}_N = \sigma \mathcal{A} / \xi$ the normal state conductance of the subsystem, and $\mathcal{G}_B = \mathcal{G}_q \mathcal{N}_{\text{max}}$ the conductance of the interface. The ratio between $\mathcal{G}_N$ and $\mathcal{G}_B$ is denoted as $r=\mathcal{G}_N/\mathcal{G}_B$ in the following. All of the above equations also hold for the propagators on the opposite side of the interface. The transition is made by interchanging the underlined and the not underlined quantities. 

Since we use coherence functions which parameterise the Green's function, the above boundary conditions should be adapted to this parameterization. One way for doing so was presented in Ref.~\cite{jacobsenCriticalTemperatureTunneling2015}, but here we choose a different approach based on the projector formalism~ \cite{shelankovDerivationQuasiclassicalEquations1985,eschrigDistributionFunctionsNonequilibrium2000}. The projectors are defined as 
\begin{equation}
    2\hat{P}_\pm = \hat{1} \pm \hat{G},
\end{equation}
allowing to rewrite  Eq.~(\ref{eq:final_boundary_cond_GF}) as
\begin{equation}
    -2r\hat{P}_\pm \pdv{\hat{P}_\pm}{z} \hat{P}_\mp = \hat{P}_\pm \hat{\overline{\mathcal{I}}} \label{eq:boundary_projector}.
\end{equation}
This equation allows to obtain 2 boundary conditions for each $\gamma$ and $\tilde{\gamma}$. The equivalence of each of the two equations as well as the equivalence to the boundary conditions calculated in Ref.~\cite{jacobsenCriticalTemperatureTunneling2015} can be shown by considering the symmetry relations implied by Eq.~(\ref{eq:boundary_projector}), i.e.,  $\hat{P}_\pm \hat{\mathcal{I}} \hat{P}_\pm = \hat{0}$. One set of boundary conditions following from Eq.~(\ref{eq:boundary_projector}) is
\begin{align}
\begin{split}
    -2r \pdv{\gamma}{z} &= \qty(\overline{\mathcal{I}}_{12} + \gamma \overline{\mathcal{I}}_{22}) (\mathit{1}-\tilde{\gamma}\gamma), \\
    -2r \pdv{\tilde{\gamma}}{z} &= \qty(\overline{\mathcal{I}}_{21} + \tilde{\gamma} \overline{\mathcal{I}}_{11}) (\mathit{1}-\gamma\tilde{\gamma}),
\end{split} \label{eq:final_boundary_cond_coherence}
\end{align}
where the particle-hole structure of the matrix current is
\begin{equation}
    \hat{\overline{\mathcal{I}}} = \begin{pmatrix} \overline{\mathcal{I}}_{11} & \overline{\mathcal{I}}_{12} \\ \overline{\mathcal{I}}_{21} & \overline{\mathcal{I}}_{22} \end{pmatrix}.
\end{equation}
Finally, we remark that the boundary conditions in Eq.~\eqref{eq:final_boundary_cond_coherence} can be used for a variety of matrix currents, i.e. the matrix current from Nazarov's boundary conditions \cite{nazarovNovelCircuitTheory1999} or the matrix current for the boundary conditions of Kuprianov and Lukichev \cite{kuprianovInfluenceBoundaryTransparency1988}.
\section{Results}\label{sec:Results}
In this section we present the results of our numerical calculations of an SC/FI/sFM/FI/SC hybrid junction described in the previous sections. Having obtained the quasiclassical Green's functions as the solutions of the Usadel equation [see Eq.~\eqref{eq:usadel_sFM}] supplemented by the boundary conditions introduced in Sec.~\ref{sec:BC}, we account for the Josephson current-phase relation in the strongly spin-polarized ferromagnet for both the charge, $I_\ch$, and the spin current, $I_\spin$. 

In what follows we discuss the Josephson current as a function of the superconducting phase difference,  $\Delta \chi = \chi_2 - \chi_1$, and the quantum-geometric phase difference, $\Delta\varphi = \varphi_2 - \varphi_2$ (the subscripts 1/2 correspond to the left/right interface), defined by the relative orientation between the magnetizations of the three ferromagnetic regions discussed below in more detail. Our main objective here is to provide a systematic theoretical study of the Josephson diode effect.

As mentioned previously, the quantity typically used to quantify this effect is the diode efficiency, defined as
\begin{equation}
    \eta_{\rm x}(\Delta\varphi) = \frac{|I_{\rm x}^+| - |I_{\rm x}^-|}{|I_{\rm x}^+| + |I_{\rm x}^-|}, \label{etadefinition}
\end{equation}
where ${\rm x} =\cc, \sn$ refers to the charge and the spin diode efficiency, respectively, and $I_{\rm x}^\pm \equiv I_{\rm x}(\Delta\chi^\pm_\cc )$. Here, $\Delta\chi^\pm_\cc $ are the superconducting phase differences at which the positive ($+$) or negative ($-$) critical charge current is reached for fixed $\Delta\varphi$, i.e., $\Delta\chi^+_\cc =\mbox{argmax}_{\Delta\chi} (I_\cc )$ and $\Delta\chi^-_\cc =\mbox{argmin}_{\Delta\chi} (I_\cc)$.
We use a definition for the spin diode efficiency that refers to the Josephson phases at the critical charge currents, as it is the charge current that usually can be controlled externally \cite{sunGatetunableSignReversal2024}. As we will show below, a geometric contribution independent of $\Delta\chi$ dominates the Fourier decomposition of the spin current in strongly asymmetric Josephson junctions of the type we consider here. In this case, a spin Josephson diode efficiency based on the extremal positive and negative spin currents, as in Ref.~\cite{maoUniversalSpinSuperconducting2024}, would yield almost everywhere 100\%, even if one of the two interfaces were entirely intransparent. The definition of Eq.~\eqref{etadefinition} we use here is directly related to the Josephson effect in both charge and spin channel, and allows us to study the diode efficiency as a function of the system's parameters, among which the quantum-geometric phase difference $\Delta\varphi$ plays a crucial role. In particular, as we qualitatively discuss in Sec.~\ref{sec:Intro} and confirm quantitatively below, the key feature leading to the Josephson diode effect is the noncoplanarity of the three magnetizations in the system $(\bm{m}_1,\bm{m}_2,\bm{M})$. Mathematically this condition is expressed as $(\bm{m}_1\times\bm{m}_2)\cdot \bm{M}\neq 0$. By assuming $\bm{M}=|\bm{M}|\bm{e}_z$ and parametrizing the misalignment of the other two magnetizations with respect to $\bm{M}$ by spherical angles [see Fig.~\ref{fig:Trilayer_sFM_sketch}], the latter relation translates into $|\bm{m}_1||\bm{m}_2||\bm{M}|\sin(\alpha_1)\sin(\alpha_2)\sin(\Delta\varphi)\neq 0$.\footnotemark[1] In other words, the noncoplanarity condition crucially depends on the relative azimuthal angle $\Delta\varphi=\varphi_2-\varphi_1$ which we term the geometric phase difference and on the polar misalignment characterized by the $\alpha_i$'s.

\subsection{Numerical procedure}
The results in this work are obtained by employing the following numerical procedure. For each point in the two-dimensional $(\Delta \chi,\Delta\varphi)$-plane we iterate the nonlinear spin-dependent Usadel equations for the quasiclassical Green's function, the self-consistency relation for the spatially dependent superconducting order parameter, as well as the nonlinear spin-dependent boundary conditions until  convergence is achieved. We then obtain the self-consistent current-phase relations, determine the global maxima and minima, and calculate the charge and spin Josephson diode efficiencies.
These last steps are greatly facilitated by a two-dimensional numerical Fourier decomposition of the current phase relations as function of $\Delta\chi$ and $\Delta\varphi$. We also analyse the current-phase relations in order to obtain spontaneous currents in a loop geometry and a switching effect between nearly 100\% spin-polarized supercurrents. Finally, we calculate the local density of states in the superconductor and in the ferromagnet.

\subsection{Current-phase relations and diode efficiencies}
In the following, we discuss the Josephson current-phase relations and the diode efficiencies, $\eta_{\rm x}$, for the cases of symmetric and asymmetric junctions. Before presenting the numerical results, we briefly discuss the scaling. As it is known from mesoscopic physics, the typical energy and lengths scales in quasiclassical superconductivity are the superconducting transition temperature, $k_B T_c$, and the coherence length $\xi=\sqrt{\hbar D/(k_BT_c)}$, respectively. On the contrary, the energy scale related to the exchange splitting, in both the sFM and the ferromagnetic interfaces, and the length scale related to the ferromagnetic interfaces are expressed in the units of the Fermi energy $E_F$ and wavelength $\lambda_F$, which are beyond the quasiclassical theory. All results for the spin-resolved currents and thus for $I_\cc$ and $I_\sn$ are expressed in the units of $I_0 = k_BT_c/(e^2R_N)$.
Here, $R_N^{-1}= \sigma_N \sqrt{\mathcal{T}_L^\uparrow \mathcal{T}_L^\downarrow} \sqrt{\mathcal{T}_R^\uparrow \mathcal{T}_R^\downarrow}$ with $\mathcal{T}_{L/R}^{\uparrow(\downarrow)}$ being the transmission probability of the left/right SC/sFM interface for the spin-$\uparrow$ or spin-$\downarrow$ channel in the case of perpendicular impact, and $\sigma_N = e^2 N_F D$ is the normal state conductivity of the superconductors.

Most of the results in this work are obtained for the temperature $T=0.1T_c$, superconducting layer's thicknesses $L_S=5\xi$, conductivity ratios $r=1$, and ferromagnetic layer's thickness $L=\xi$. The magnetization direction of the ferromagnet $\bm{M} \parallel \bm{e}_z$ and the polar angle characterizing the left and right interfaces' magnetizations $\alpha_L=\alpha_R=\pi/2$ [see Fig.~\ref{fig:Trilayer_sFM_sketch}], i.e., $\bm{m}_1$ and $\bm{m}_2$ lie in the $x-y$ plane. To get more insights into the role of these parameters we refer to Appendix~\ref{sec:appendix:parameter_discussion}. 

\begin{figure}[t]
    \centering
    \includegraphics[width=\linewidth]{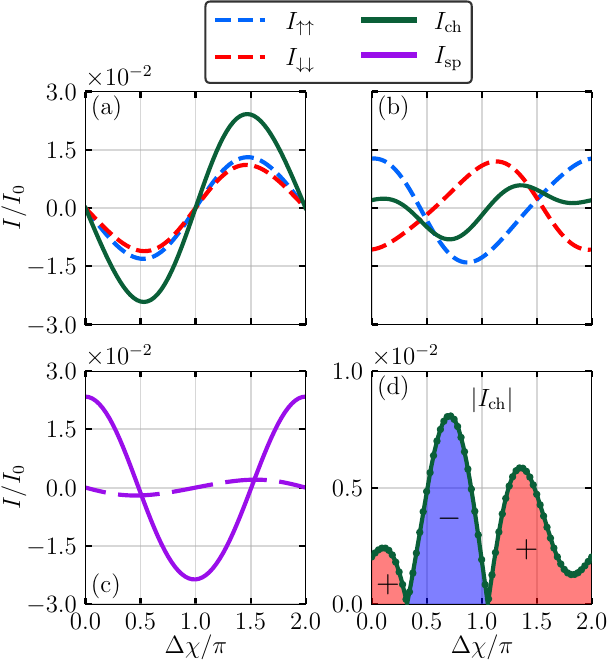}
    \caption{Spin-resolved current-phase relation (CPR) $I_{\upup}(\Delta \chi)$, $I_{\dodo}(\Delta\chi)$, and charge current $I_\cc (\Delta\chi )$ for (a) a coplanar arrangement of $(\bm{m}_1,\bm{m}_2,\bm{M})$, i.e., $\Delta\varphi = 0$, and (b) a noncoplanar arrangement,
    here $\Delta\varphi/\pi = 0.44$. (c) spin currents $I_{\sn}(\Delta\chi)$ corresponding to panel (a) and (b)
    for the coplanar case (dashed line) and the noncoplanar case (solid line).  (d) absolute value of $I_{\cc}$ from panel (b), with the sign denoted by red ($+$) and blue ($-$) color; dots represent values calculated via QCGFs and used for the Fourier decomposition of the CPR.
    Parameters are appropriate for a symmetric junction with $V_\mathrm{B} = 1.3 E_F$, $V_\mathrm{sFM} = 0$, $J_\mathrm{sFM} = J_\mathrm{B} = 0.4 E_F$, $d_\uparrow = 0.6 \lambda_F / 2\pi$ and $d_\downarrow = 0.8 \lambda_F / 2\pi$. The left and the right interface differ only in $\varphi $.}
    \label{fig:CPR_diode_effect_symm}
\end{figure}

We first discuss the Josephson effect occurring in the case of coplanar magnetizations and symmetric interfaces. Such a configuration is realized for the cases $\sin(\Delta\varphi) = 0\implies \Delta\varphi=k\pi,  k \in \mathbb{Z}$. The resulting current-phase relations calculated for $\Delta\varphi=0$ are shown in Fig.~\ref{fig:CPR_diode_effect_symm}(a), where the dashed lines represent the spin-resolved currents whereas the solid green line represents the charge current. The system's parameters are shown in the caption. The corresponding spin current is represented by the dashed line in Fig.~\ref{fig:CPR_diode_effect_symm}(c) and it is well-approximated by the standard sinusoidal shape. Note that in this case, all currents exhibit the normal Josephson effect, $I(\Delta\chi)=-I(-\Delta\chi)$.

However, the situation of nonvanishing $\Delta\varphi$, i.e., a noncoplanar configuration, drastically modifies the Josephson current-phase relation as shown in Fig.~\ref{fig:CPR_diode_effect_symm}(b). The curves are calculated for $\Delta\varphi=0.44\pi$ and other parameters being the same as in Fig.~\ref{fig:CPR_diode_effect_symm}(a). There are two new phenomena compared to the coplanar case. First, the anomalous Josephson effect, characterized by $I(\Delta\chi=0)\neq0 \implies I(\Delta\chi)\neq-I(-\Delta\chi)$, occurs. Second, the presence of higher harmonics gives rise to the Josephson diode effect for the charge current (green solid line). This effect is explicitly visible in Fig.~\ref{fig:CPR_diode_effect_symm}(d) where the absolute value of the charge current from (b) is plotted. The spin current [solid line in Fig.~\ref{fig:CPR_diode_effect_symm}(c)] displays a current-phase relation strongly shifted in $\Delta\chi$ compared to the coplanar case  [dashed line in Fig.~\ref{fig:CPR_diode_effect_symm}(c)].
A detailed analysis of the diode efficiencies and the harmonic decomposition of the current-phase relation for charge and spin currents is provided below.

\begin{figure}
    \centering
    \includegraphics[width=\linewidth]{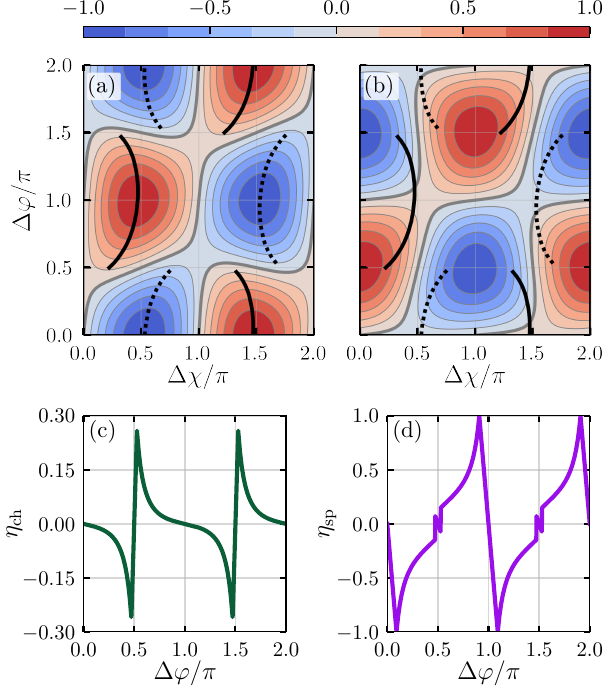}
    \caption{Functional dependence of  (a) charge, $I_\cc$, and (b) spin current, $I_\sn$, of a symmetric junction on the superconducting phase difference $\Delta\chi$ and the geometric phase $\Delta\varphi$. Currents in each panel are normalized to their maximal values.
 Solid black and dashed lines show the branches of $\Delta\chi^+(\Delta\varphi)$ and $\Delta\chi^-(\Delta\varphi)$ [see the text for definition], respectively. The thick grey line in (a) and (b) denotes zero current. In (c) and (d) the functional dependence of the charge and spin diode efficiency on $\Delta\varphi$ is shown. The potentials and exchange fields of the ferromagnetic insulating layers and metallic ferromagnet are the same as in Fig. \ref{fig:CPR_diode_effect_symm}.}
    \label{fig:colormap_ch_sp_curr_and_spon_curr_diode_eff_SYMM}
\end{figure}

To obtain a better insight in the role of the \mbox{quantum-geometric phase difference}, we consider the charge and the spin current (quantified by $I_\cc = I_\upup + I_\dodo$, $I_\sn = I_\upup - I_\dodo$) as a function of $\Delta\chi$ and $\Delta\varphi$, shown in Figs.~\ref{fig:colormap_ch_sp_curr_and_spon_curr_diode_eff_SYMM}(a) and \ref{fig:colormap_ch_sp_curr_and_spon_curr_diode_eff_SYMM}(b), respectively. Both panels are plotted for the same selection of parameters as in Fig.~\ref{fig:CPR_diode_effect_symm}. 
Apparently, $I_\cc(\Delta\chi)$ has significant contributions from higher harmonics except for  $\Delta\varphi\approx k\pi,\, k\in\mathbb{Z}$ where the first harmonic is dominant, i.e., $I_\cc \approx I^\pm_\cc(\Delta\varphi)\sin(\Delta\chi)$. In addition, we observe $\Delta\varphi$-driven $0-\pi$ transitions. The spin current, shown in panel (b), is for most values of $\Delta\varphi$ well-approximated by $I_\sn \sim \pm \cos(\Delta\chi)$. 
 Panels (c) and (d) show the charge $\eta_\cc$ (green) and the spin $\eta_\sn$ (violet) diode efficiency, respectively, as a function of $\Delta\varphi$, obtained from a numerical Fourier decomposition of the currents shown in panels (a) and (b).
 We find a significant diode effect in both the charge and the spin current, and obtain a perfect spin diode for distinct $\Delta\varphi$'s, i.e., $\eta_\sn = \pm 1$. The diode effect vanishes for $\Delta\varphi=k\pi$, which corresponds to the case of a coplanar spin arrangement. The diode effect vanishes also for $\Delta\varphi=\frac{\pi}{2}+k\pi $. This can be explained by considering the phase-factors acquired by $\upup$ and $\dodo$ correlations which either coincide for $\Delta\varphi=k\pi$ or differ by multiplies of $\pi$ for $\Delta\varphi=\frac{\pi}{2}+k\pi $.
Also, $\eta_\cc$ tends to peak around $\Delta\varphi=\frac{\pi}{2}+k\pi$, whereas perfect efficiency in $\eta_\sn$ can occur for any $\Delta\varphi$ differing from $k\frac{\pi}{2}$, as shown in Appendix~\ref{sec:appendix:parameter_discussion}. In
Fig.~\ref{fig:colormap_ch_sp_curr_and_spon_curr_diode_eff_SYMM} it happens to be close to $\Delta\varphi=k\pi$.

Changing the parameters characterising the strong spin-splitting yields a change in the magnitude and the explicit functional dependencies, however the effect itself is present as long as the FI layers and the sFM are spin-active, and the sFM has two itinerant spin bands (i.e., is not half-metallic) with different densities of states. For a discussion of the effect over a wide range of parameters see Appendix \ref{sec:appendix:parameter_discussion}.

\begin{figure}[t]
    \centering
    \includegraphics[width=\linewidth]{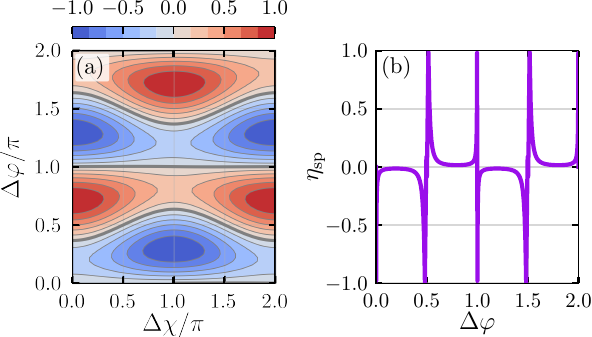}
    \caption{(a) Spin current, $I_\sn$, and (b) spin diode efficiency, $\eta_\sn$, as in Fig. \ref{fig:colormap_ch_sp_curr_and_spon_curr_diode_eff_SYMM}(b) and (d), however for an asymmetric junction where the left FI layer is the same as in the symmetric case (see Fig.~\ref{fig:CPR_diode_effect_symm}) and the right FI layer's width is the same for both spin projections, $d_R = 4.16 \lambda_F / 2\pi$. The remaining parameters are the same as in Fig. \ref{fig:CPR_diode_effect_symm}.}
    \label{fig:colormap_sp_curr_diode_eff_ASYMM}
\end{figure}

In the previous discussion, we assumed that both interfaces in the system are identical up to different $\varphi$'s. We now briefly discuss the case of different widths of the ferromagnetic insulating layers, i.e., different transmission probabilities on either side of the sFM. For definiteness, we use the same parameters as in
Fig.~\ref{fig:CPR_diode_effect_symm} except that now the right FI layer is less transparent having the width $d_R=4.16\lambda_F/2\pi$ (for both spin projections). Figure~\ref{fig:colormap_sp_curr_diode_eff_ASYMM}(a) shows the spin current $I_\sn$ as a function of the superconducting, $\Delta\chi$, and the geometric phase difference, $\Delta\varphi$, whereas Fig.~\ref{fig:colormap_sp_curr_diode_eff_ASYMM}(b) shows the corresponding spin diode efficiency as function of $\Delta\varphi$.
In this case, higher harmonics of the charge current are strongly suppressed and the Josephson current-phase relation is approximately sinusoidal. 
 On the other hand, the spin current $I_\sn(\Delta\chi,\Delta\varphi)$ does not change its sign as a function of $\Delta\chi$ in a wide range of $\Delta\varphi$. We observe sign changes with $\Delta\chi$ only in relatively narrow ranges of $\Delta\varphi$ close to $\Delta\varphi \approx \frac{\pi}{2}+k\pi$. The spin Josephson diode efficiency, shown in Fig.~\ref{fig:colormap_sp_curr_diode_eff_ASYMM}(b), exhibits the perfect efficiency only close to $\Delta\varphi \approx \frac{\pi}{2}+k\pi$, and extremely close to $\Delta\varphi \approx k\pi$. A more detailed discussion on the cross-over from symmetric to strongly asymmetric junctions can be found in Appendix \ref{sec:appendix:parameter_discussion}.

\begin{figure*}[t!]
    \centering
    \includegraphics[width=0.95\textwidth]{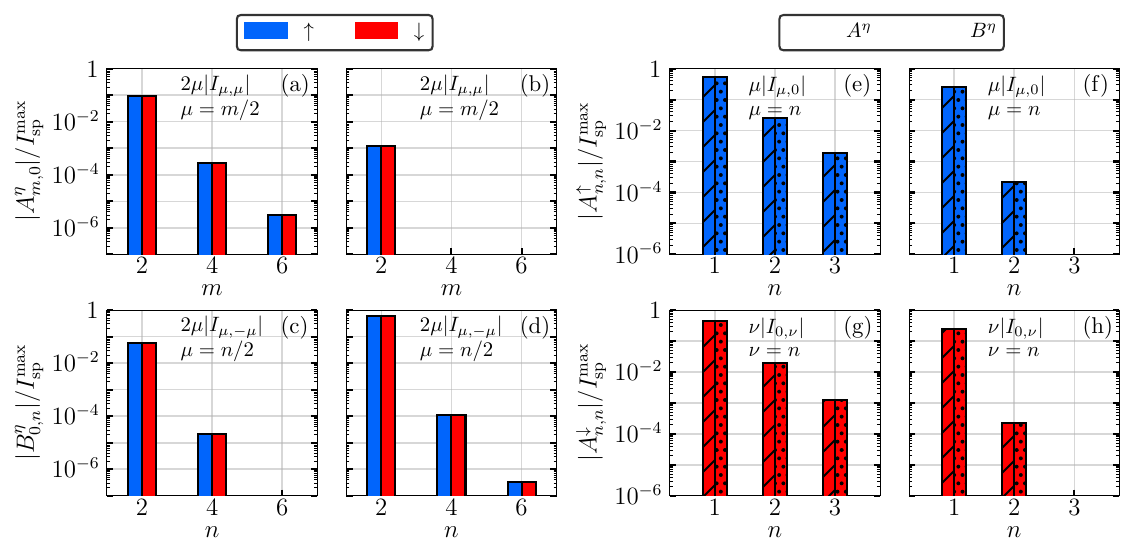}
    \caption{Fourier components of the spin-resolved currents corresponding to the charge and spin currents shown in Figs.~\ref{fig:colormap_ch_sp_curr_and_spon_curr_diode_eff_SYMM}(a), \ref{fig:colormap_ch_sp_curr_and_spon_curr_diode_eff_SYMM}(b) and \ref{fig:colormap_sp_curr_diode_eff_ASYMM}(a). The panels (a),(c),(e), and (g) of this figure show the results for the symmetric case [see Fig.~\ref{fig:colormap_ch_sp_curr_and_spon_curr_diode_eff_SYMM}], whereas the panels (b),(d),(f), and (h) shows the results for the asymmetric junction case [see Fig.~\ref{fig:colormap_sp_curr_diode_eff_ASYMM}]. The different colors indicate the two spin-bands for which we perform the Fourier analysis, $\uparrow$ in blue and $\downarrow$ in red. In (a) and (b) we plot the absolute value of the $A_{m,0}^\eta$ component and in panels (c) and (d) the absolute values of the $B_{0,n}^\eta$ components are shown. Panels (e) - (h) show the absolute values of the Fourier components with equal contributions in $\Delta\chi$ and $\Delta\varphi$, i.e. $A_{n,n}^\eta$ (dashed bar) and $B_{n,n}^\eta$ (dotted bar), for the $\uparrow$-band in panels (e) and (f) and for the $\downarrow$-band in panels (g) and (h). All Fourier components are normalized to the maximum spin current for the corresponding case, i.e. $I^\mathrm{max}_\sn = \mathrm{max}_{(\Delta\chi,\Delta\varphi)} I_\sn(\Delta\chi,\Delta\varphi)$.}
    \label{fig:Fourier_comp}
\end{figure*}

\subsection{Fourier decomposition}
As already indicated, the presence of higher harmonics in the Josephson current-phase relation is crucial for the Josephson diode effect. To get more insight, we present a detailed Fourier analysis of the current-phase relation both in $\Delta\chi$ and $\Delta\varphi$. Since both the charge, $I_\cc$, and the spin current, $I_\sn$, are mediated by equal-spin triplet contributions, $I_{\eta\eta}, \eta=\uparrow,\downarrow$, in this subsection we focus on them separately. 
We will show below that the current phase relations can be written in the form 
\begingroup
\allowdisplaybreaks
\begin{align}
    I_{\uparrow\uparrow} &= \frac{1}{2}\sum_{\mu,\nu=-\infty}^\infty  (-1)^{\mu+\nu} \mu I_{\mu,\nu} \sin\psi_{\mu,\nu}, \label{eq:I_uu} \\ 
    I_{\downarrow\downarrow} &= \frac{1}{2}\sum_{\mu,\nu=-\infty}^\infty  (-1)^{\mu+\nu} \nu I_{\mu,\nu} \sin\psi_{\mu,\nu}, \label{eq:I_dd}
\end{align}
\endgroup
where $I_{-\mu,-\nu}=I_{\mu,\nu}$, and 
\begin{equation}
\psi_{\mu,\nu} = (\mu+\nu)\Delta\chi - (\mu-\nu)\Delta\varphi \label{effJos}
\end{equation}
is an \textit{effective Josephson phase}. This form has been suggested for ballistic systems \cite{eschrigSpinpolarizedSupercurrentsSpintronics2015,eschrigPhasesensitiveInterfaceProximity2019}, and we show here that it also holds for diffusive systems.
Eqs.~\eqref{eq:I_uu}-\eqref{effJos} have an appealing physical interpretation in terms of transferred equal-spin pairs across the junction \cite{eschrigSpinpolarizedSupercurrentsSpintronics2015}. They present the current-phase relation for the case of a coherent transfer of $\mu$ $\uparrow\uparrow$ and $\nu$ $\downarrow\downarrow$ equal-spin pairs (positive values denote transmissions in positive current direction, negative values denote transmissions in the opposite direction). Taking into account that the ground state for $\Delta \varphi=0$ is a $\pi$-junction, the Josephson phase acquired for a coherent transfer of a single
 $\uparrow\uparrow/\downarrow\downarrow$ pair is $\Delta\chi \pm \Delta\varphi + \pi$. Thus, for the coherent transport of $\mu$ $\uparrow\uparrow$-pairs and $\nu$ $\downarrow\downarrow$-pairs a Josephson phase $\mu(\Delta\chi - \Delta\varphi + \pi) + \nu(\Delta\chi + \Delta\varphi + \pi) = (\mu+\nu)\Delta\chi - (\mu-\nu)\Delta\varphi + (\mu+\nu) \pi$ appears, leading to Eqs.~\eqref{eq:I_uu}-\eqref{effJos}.
Note that within the system at hand only even numbers of electrons are transmitted through each spin band of the the sFM, as short-range pair amplitudes, involving both bands, are negligible for strong spin polarization.

We now proceed to discuss a number of symmetry relations for a general Fourier series expansion that allow us to derive Eqs.~\eqref{eq:I_uu}-\eqref{effJos} for our system on general grounds. Using the general QCGF technique we numerically confirm these symmetry relations. Details are given in Appendix \ref{sec:appendix:Fourier}.
We make use of the fact that $I_{\eta\eta}$ under the time reversal symmetry transforms as follows \cite{eschrigSpinpolarizedSupercurrentsSpintronics2015}:
\begin{equation}
 I_{\eta\eta}(\Delta\chi,\Delta\varphi) = - I_{\eta\eta}(-\Delta\chi,-\Delta\varphi).
 \label{symmIetaeta}
\end{equation}
This relation is derived by noting that the angle $\Delta \varphi $ is defined by three magnetic vectors that behave odd under time reversal. We also chose as spin quantization axis  the exchange field in the sFM, and therefore both  behave odd under time reversal, allowing for the derivation of the above relation for majority and minority spin bands.
A general Fourier series expansion compatible with Eq.~\eqref{symmIetaeta}  is given by (again, $\eta \in \{\uparrow, \downarrow\}$)
\begin{align}
        I_{\eta\eta} =& \frac{1}{2}\sum_{m=1}^\infty 
        A_{m,0}^\eta \sin(m\Delta\chi) +
        \frac{1}{2}\sum_{n=1}^\infty 
        B_{0,n}^\eta \sin(n\Delta\varphi) \nonumber \\
        &+\sum_{m,n=1}^\infty \Big[A_{m,n}^\eta \sin(m\Delta\chi)\cos(n\Delta\varphi) \label{eq:general_Fourier_ansatz}\\
        &\qquad \qquad +\,  B_{m,n}^\eta \cos(m\Delta\chi)\sin(n\Delta\varphi)\Big]. \nonumber
\end{align}
In Appendix~\ref{sec:appendix:Fourier} we list symmetry relations between the coefficients $A_{m,n}^\eta$, $B_{m,n}^\eta$ which are confirmed by our numerical calculations, and which render Eq.~\eqref{eq:general_Fourier_ansatz} equivalent to Eqs.~\eqref{eq:I_uu}-\eqref{effJos}. In particular, we find that coefficients are only nonzero when either both $m$ and $n$ are even or both $m$ and $n$ are odd (i.e., when $m\pm n$ is even). Given this fact, the coefficients relating Eq.~\eqref{eq:I_uu} to Eq.~\eqref{eq:I_dd} are
\begin{align}
   (-1)^{\mu+\nu} |\mu| I_{\mu,\nu}=
    \frac{A^\uparrow_{|\mu+\nu|,|\mu-\nu|}-B^\uparrow_{|\mu+\nu|,|\mu-\nu|}}{2},\\
    (-1)^{\mu+\nu} |\nu| I_{\mu,\nu}=
    \frac{A^\downarrow_{|\mu+\nu|,|\mu-\nu|}-B^\downarrow_{|\mu+\nu|,|\mu-\nu|}}{2},
\end{align}
and we define formally $A^\eta_{0,2\nu}=0$ and $B^\eta_{2\mu,0}=0$.
In the following, we focus on the discussion of the most relevant Fourier contributions $A^\eta_{m,0}$, $B^\eta_{0,n}$, $A_{n,n}^\eta$, and  $B_{n,n}^\eta$ which are presented in Fig.~\ref{fig:Fourier_comp}, and which correspond to $I_{\mu,\mu}$, $I_{\mu,-\mu}$, $I_{\mu,0}$, and $I_{0,\nu}$, correspondingly.

We start by considering the absolute values of $A^\eta_{m,0}$ [see Figs.~\ref{fig:Fourier_comp}(a) and \ref{fig:Fourier_comp}(b)] and $B^\eta_{0,n}$ [see Figs.~\ref{fig:Fourier_comp}(c) and \ref{fig:Fourier_comp}(d)]. Note that the left column in both cases corresponds to the symmetric junction from Fig.~\ref{fig:colormap_ch_sp_curr_and_spon_curr_diode_eff_SYMM} whereas the right one refers to the asymmetric junction considered in Fig.~\ref{fig:colormap_sp_curr_diode_eff_ASYMM}. It is evident that in both cases the Fourier components for the $\uparrow$ (red bars) and the $\downarrow$ channel (blue bars) coincide. Furthermore, we obtain a sign change for $B_{0,n}^\eta$ (which is not explicitly visible here since we show the absolute values for better comparison). Thus, the symmetries 
of these contributions
can be summarized as 
\begin{equation}
    A_{m,0}^\uparrow =  A_{m,0}^\downarrow \quad \text{and} \quad B_{0,n}^\uparrow = -B_{0,n}^\downarrow.
\end{equation}
Only Fourier coefficients of even order are nonzero.
The physical interpretation of $\frac{1}{2} A^\eta_{2\mu,0}=\mu I_{\mu,\mu}$ is in terms of crossed pair transmission processes, where in both spin bands $\mu $ equal-spin pairs are transferred across the junction in the same direction. These terms are therefore equal for $\eta=\uparrow,\downarrow$ and contribute only to the charge current. On the other hand, $\frac{1}{2} B^\uparrow_{0,2\nu}=\nu I_{\nu,-\nu}$ represent processes where $\nu $ equal-spin pairs are transferred in opposite direction for the two spin bands. These terms have opposite signs for opposite spin and therefore contribute only to the spin current.
Considering the amplitudes of the Fourier components, which are normalized to the respective maximum spin current, shows that in the symmetric case [see Figs.~\ref{fig:Fourier_comp}(a) and~\ref{fig:Fourier_comp}(c)] crossed pair transmission processes of an equal number of pairs in the two spin bands in the same direction dominate, whereas 
for the asymmetric case [see Figs.~\ref{fig:Fourier_comp}(b) and~\ref{fig:Fourier_comp}(d)] the transmission processes  of an equal number of pairs in the two spin bands in opposite directions dominate.

Next, we discuss the Fourier components with equal contributions from $\Delta\chi$ and $\Delta\varphi$, i.e., $A_{n,n}^\eta$ and $B_{n,n}^\eta$. 
The absolute values of them are plotted in Figs.~\ref{fig:Fourier_comp}(e) and~\ref{fig:Fourier_comp}(f) ($\uparrow$ spin band; blue) and Figs.~\ref{fig:Fourier_comp}(g) and~\ref{fig:Fourier_comp}(h) ($\downarrow$ spin band; red). Different patterns corresponds to different Fourier coefficients as referred in the legend. As in panels (a)-(d) discussed above here the left column [Figs.~\ref{fig:Fourier_comp}(e) and~\ref{fig:Fourier_comp}(g)] corresponds to the symmetric and the right one [Figs.~\ref{fig:Fourier_comp}(f) and~\ref{fig:Fourier_comp}(h)] to the asymmetric junction. We obtain that these contributions are also constrained by symmetry relations which are
\begin{equation}
    A_{n,n}^\uparrow = -B_{n,n}^\uparrow \quad \mathrm{and} \quad A_{n,n}^\downarrow = B_{n,n}^\downarrow. \label{eq:fourier:symmetry_diagonal}
\end{equation}
Note that we again plot the absolute values for better comparison. The physical interpretation of these terms is that they correspond to processes where $n$ pairs are transmitted in one spin band, and zero in the other spin band. For $\mu $ pairs transmitted only in the spin-$\uparrow$ band, $A^\uparrow_{\mu,\mu}=(-1)^\mu \mu I_{\mu,0}$, and for $\nu $ pairs transmitted only in the spin-$\downarrow$ band, $A^\downarrow_{\nu,\nu}=(-1)^\nu \nu I_{0,\nu}$.

Moreover, we see that in both cases the components exponentially decrease with increasing order $n$. Comparing the magnitudes of the various terms, we find that for symmetric junctions a typical hierarchy $|I_{1,0}+I_{0,1}|\gg |I_{1,0}-I_{0,1}|\sim 2I_{1,1}\sim 2I_{1,-1}\sim 2|I_{2,0}+I_{0,2}|\gg 2|I_{2,0}-I_{0,2}|$ is present.
For strongly asymmetric junctions, in contrast, $|I_{1,0}+I_{0,1}|\sim 2I_{1,-1} \gg  |I_{1,0}-I_{0,1}|\gtrsim 2I_{1,1}\gtrsim 2|I_{2,0}+I_{0,2}| \gg 2|I_{2,0}-I_{0,2}|$ typically holds.

Finally, the $A_{m,n}^\eta$'s and $B_{m,n}^\eta$'s for $m,n>0$ and $m\neq n$ have a quite small magnitude and we do not discuss them here as they do not give additional insights. However, we find that these contributions also vanish for $(m\pm n)$ odd. The consequence of this finding is, that of all processes where a fixed number of pairs are transferred across both spin bands, those with equal number of transferred pairs in the two spin bands dominate over those where one spin band carries more pairs than the other (for example, $|I_{1,1}|>|I_{2,0}|$, $|I_{2,2}|>|I_{3,1}|, |I_{4,0}|$ etc). This can be understood in terms of the fact, that when spin-triplet amplitudes are present in the sFM, an equal number of equal-spin pairs can recombine into spin-singlet Cooper pairs on the SC side of the interface without resorting to the triplet rotation mechanism, whereas for any surplus pairs the triplet rotation mechanism is necessary in order to transform them into singlet Cooper pairs.

In summary, we find that within numerical accuracy our system is described by Eqs.~\eqref{eq:I_uu}-\eqref{effJos}.

\subsection{Analytical model} \label{sec:analytic_model}

It follows from Fig.~\ref{fig:Fourier_comp} that the lowest harmonic contributions dominate in all cases. We find that a model that includes terms up to second order in the number of transferred pairs, i.e., $\abs{\mu}+ \abs{\nu}\leq 2 $, 
describes the numerical data very closely. Therefore, we write
\begin{align}
\begin{split}
    I_{\uparrow\uparrow} &\approx I_{1,1}\sin\qty(2\Delta\chi) - I_{1,-1}\sin\qty(2\Delta\varphi) \\
    &- I_{1,0} \sin\qty(\Delta\chi - \Delta\varphi)+ 2I_{2,0} \sin\qty(2\Delta\chi - 2\Delta\varphi)
    , \label{eq:CPR_approx_upup}
\end{split}
\\
\begin{split}
    I_{\downarrow\downarrow} &\approx I_{1,1}\sin\qty(2\Delta\chi) + I_{1,-1}\sin\qty(2\Delta\varphi) \\
    & - I_{0,1} \sin\qty(\Delta\chi + \Delta\varphi)+ 2I_{0,2} \sin\qty(2\Delta\chi + 2\Delta\varphi)
    . \label{eq:CPR_approx_dodo}
\end{split}
\end{align}
This yields the following expressions for $I_\cc = I_\upup + I_\dodo$ and  $I_\sn = I_\upup - I_\dodo$
\begingroup
\allowdisplaybreaks
\begin{align}
\begin{split}
    I_\cc &\approx 2 I_{1,1} \sin\qty(2\Delta\chi) \\
    &-  I_{1+} \sin\qty(\Delta\chi) \cos\qty(\Delta\varphi)
    + I_{1-}\cos\qty(\Delta\chi)\sin\qty(\Delta\varphi)\\
    &+I_{2+} \sin\qty(2\Delta\chi) \cos\qty(2\Delta\varphi)
    -I_{2-}\cos\qty(2\Delta\chi)\sin\qty(2\Delta\varphi),
    \label{Icappr}
\end{split} \\
\begin{split}
    I_\sn &\approx -2 I_{1,-1} \sin\qty(2\Delta\varphi)
    \\
    &- I_{1-} \sin\qty(\Delta\chi) \cos\qty(\Delta\varphi) 
     +I_{1+}\cos\qty(\Delta\chi)\sin\qty(\Delta\varphi)\\
     &+ I_{2-} \sin\qty(2\Delta\chi) \cos\qty(2\Delta\varphi) 
     -I_{2+}\cos\qty(2\Delta\chi)\sin\qty(2\Delta\varphi), 
     \label{Isappr}
\end{split}
\end{align}
\endgroup
with the definition
\begin{align}
   I_{n\pm}=n(I_{n,0} \pm I_{0,n}).
\end{align}

In order to derive analytic expressions for the maximal Josephson diode efficiency, we write $\Delta\varphi=\frac{\pi}{2}+\delta $ and expand Eqs.~\eqref{Icappr}-\eqref{Isappr} for small $\delta$. Using $\cos(\Delta \varphi)\approx -\delta$, $\sin (\Delta\varphi)\approx 1$, $\cos(2\Delta\varphi)\approx -1$, $\sin(2\Delta\varphi)\approx -2\delta $, this leads to the following model CPR:
\begin{align}
    I_\cc &\approx 2I_{1,1} \sin (2\Delta\chi) +\delta \cdot I_{1+} \sin(\Delta\chi) + I_{1-} \cos(\Delta\chi) \nonumber \\ & \qquad\qquad - I_{2+}\sin(2\Delta\chi)+2\delta \cdot I_{2-} \cos(2\Delta\chi).
    \label{AnIch1}
\end{align}
In our expansion, we will consider $I_{2+}/I_{1+}$, $I_{1,1}/I_{1+}$, and $I_{n-}/I_{n+}$ to be of the same order of magnitude as $\delta $. This is the case close to the maximal charge Josephson diode efficiency, where $\delta $ is small. We then can neglect the term proportional to $\delta \cdot I_{2-}$. 
We define the two parameters
\begin{align}
A=2I_{1,1} -I_{2+},\qquad B=I_{1-} \label{AB} .
\end{align}
For $|B|<\sqrt{8}|A|$, the Josephson diode efficiency has maximal magnitude at $\delta=\pm \frac{I_{1-}}{I_{1+}}$, and the model CPR for this value of $\delta $ is given by
\begin{align}
    I_\cc &\approx A\sin (2\Delta\chi) + B \left[ \cos(\Delta\chi) \pm \sin(\Delta\chi) \right] .
    \label{CPRchargeefficiency}
\end{align}
This expression has for $|B|<\sqrt{8}|A|$ four extrema. For example, for $\delta = -\frac{I_{1-}}{I_{1+}}$, $A>0$ they are
\begin{alignat}{4}
    & \Delta \chi_1 && =\arctan \left(\frac{S-B}{S+B}\right), && 
    ~I_{\cc,1} &&=A+\frac{B^2}{4A},\\
    & \Delta \chi_2 && =\frac{3}{4}\pi, && 
    ~I_{\cc,2} &&=-A-\sqrt{2}B,\\
    & \Delta\chi_3 && =\pi+\arctan \left(\frac{S+B}{S-B}\right), && 
    ~I_{\cc,3} &&=I_{c}(\Delta\chi_1),\\
    & \Delta\chi_4 && =2\pi-\frac{\pi}{4}, && 
    ~I_{\cc,4} &&=-A+\sqrt{2}B .
\end{alignat}
with $S=\sqrt{8A^2-B^2}$, and $\Delta \chi_n$ are given modulo $2\pi$. For $\delta $ deviating from the optimal value $\pm \frac{I_{1-}}{I_{1+}}$, the degeneracy for the critical currents at $\Delta\chi_1$ and $\Delta\chi_3$ is lifted.
The critical currents are $I_\cc^+=A+\frac{B^2}{4A}$, $I_\cc^-=-A-\sqrt{2}|B|$. For $A<0$ a similar consideration gives $I_\cc^+=|A|+\sqrt{2}|B|$, $I_\cc^-=-|A|-\frac{B^2}{4|A|}$. The results for $\delta=+\frac{I_{1-}}{I_{1+}}$ are obtained by $\Delta \chi_n\to -\Delta \chi_n$ mod $2\pi$ and $A\to -A$, i.e.,
the same critical currents result with opposite signs. Collecting all together, the charge Josephson diode efficiency with the largest magnitude, $\widehat \eta_\cc $, is obtained as 
\begin{align}
    \widehat\eta_\cc\approx\pm \frac{4\sqrt{2}|AB|-B^2}{(\sqrt{8}|A|+|B|)^2}\mbox{sign}(A),
    \label{etamodel1}
\end{align}
where the $\pm $ sign in front of the expression corresponds to $\delta=\pm \frac{I_{1-}}{I_{1+}}$.
Comparing this expression with the numerical results of Fig.~\ref{fig:colormap_ch_sp_curr_and_spon_curr_diode_eff_SYMM}, we obtain from Eq.~\eqref{etamodel1} that $\abs{\widehat\eta_\cc} \approx 26 \%$, which is in excellent agreement with the numerical value. The maximal efficiency that can be obtained from Eq.~\eqref{etamodel1} is for $|B|=\sqrt{2}|A|$ and is 33.$\overline{3}$\%.  The same consideration can be made for $\Delta \varphi$ near $(k+\frac{1}{2})\pi$ with integer $k$, and
the  value of $\Delta \varphi$ for optimal efficiency is then given approximately by 
\begin{align}
   \widehat{\Delta\varphi}_\cc&\approx \left(k+\frac{1}{2}\right) \pi\pm \frac{I_{1,0}-I_{0,1}}{I_{1,0}+I_{0,1}}
\end{align}
with $k$ integer.

For $|B|>\sqrt{8}|A|$, Eq.~\eqref{CPRchargeefficiency} exhibits only two extrema. In this case, $I_\cc^+=-A+\sqrt{2}|B|$ and $I_\cc^-=-A-\sqrt{2}|B|$. The corresponding values of the Josephson diode efficiency are then
\begin{align}
    \eta_\cc \approx \pm\frac{A}{\sqrt{2}|B|}.
    \label{etamodel2}
\end{align}
Here, the maximal efficiency in the range of applicability is attained for $|B|=\sqrt{8}|A|$, for which the values of Eqs.~\eqref{etamodel1} and \eqref{etamodel2} coincide, and is 25\%. However, the optimal value $\widehat\eta_\cc $ is slightly higher than this value, because in this parameter range $\delta$ for the optimal efficiency deviates from $\pm \frac{I_{1-}}{I_{1+}}$ (it assumes a smaller magnitude).

It is clear, that these expressions for the charge diode efficiency require $A$ and $B$ to be nonzero for a JDE to occur. As $B=I_{1-}$, this means $I_{1,0}\ne I_{0,1}$. Also, taking into account that $|I_{1,1}|>|I_{2,0}|,|I_{0,2}|$, the condition $I_{1,1}\ne 0$ must be fulfilled. Finally, the JDE is absent for $\delta =0$, i.e., for $\Delta\varphi=\frac{\pi}{2}+k\pi $ with $k$ integer.

We now turn to the spin current. In particular, we will show, that 
for certain values of $\Delta\varphi$ the spin current efficiency can reach values of 100\%. 
In general, as long as $\cos(\Delta \varphi)$ does not become too small, the charge current, Eq.~\eqref{Icappr}, is dominated by the term $-I_{1+}\cos(\Delta\varphi)\sin(\Delta \chi)$.  
The phases for the critical charge currents are then close to $\pm \frac{\pi}{2}$, and given by 
\begin{align}
    \Delta \chi_{1/2}\approx\pm\frac{\pi}{2} \pm \frac{4I_{1,1}+2I_{2+}\cos(2\Delta\varphi) \pm I_{1-}\sin(\Delta\varphi)}{I_{1+}\cos(\Delta \varphi)}
    \label{chi12}
\end{align}
and the corresponding spin currents are then 
\begin{align}
    I_{\sn ,1/2}\approx \frac{\mp B-2\sin(\Delta\varphi) \left\{ A+C\cos^2(\Delta\varphi) \right\}}{\cos(\Delta\varphi )}
    \label{Is12}
\end{align}
with $A$ and $B$ given in Eq.~\eqref{AB}, and 
\begin{align}
    C=2I_{1,-1}+I_{2+}. \label{eq:def_C}
\end{align}
Equation~\eqref{Is12} holds under the condition that $\Delta \chi_{1,2}$ are close to $\pm \frac{\pi}{2}$.
The condition, that the spin current vanishes for one of those phases (and therefore a spin diode efficiency of 100\% occurs) reads
\begin{align}
    \pm B =-2\sin(\Delta\varphi) \left\{A+C\cos^2(\Delta\varphi)\right\}.
    \label{condition}
\end{align}
For this equation to have a solution, $|B|$ has to be sufficiently small, at least smaller than max($2|A|$,$4|C|$). 
For $|B^2C|\ll 4|A+C|^3$, we obtain $\widehat\eta_\sn =\pm 1$ for 
\begin{align}
    \widehat{\Delta \varphi}_\sn \approx k\pi \mp \arcsin\left( \frac{2B}{4(A+C)-B^2C/(A+C)^2}\right). \label{eq:phi_sp}
\end{align}
In this expression, the argument of the $\arcsin $ must have a magnitude $< 1$. In the limit $|C|\ll |A|$, appropriate for symmetric junctions, this leads to the restriction $|B|<2|A|$. On the other hand, for $|C|\gg |A|$, appropriate for strongly asymmetric junctions, the restriction reads $|B+C|<\sqrt{5}|C|$.
For $|C|\gg|A|,|B|$ there is another solution of Eq.~\eqref{condition}, which results from a small $\cos^2(\Delta\varphi)$. In this case, also appropriate for strongly asymmetric junctions, a spin diode efficiency of 100\% is achieved for (assuming $C>0$)
\begin{align}
\widehat{\Delta \varphi}_{\sn } \approx \frac{\pi}{2}+k\pi \pm \arcsin \sqrt{\frac{2(B-2A)}{4C-B}}.
\label{aroundpihalf}
\end{align}
under the condition that $B>2A$. To ensure that the approximations made in Eq.~\eqref{chi12} still hold, the condition $|A|,|B|\ll I_{1+}\sqrt{(B-2A)/2C}$ should also apply.

The right hand side of Eq.~\eqref{condition} is a third order polynomial in $\sin(\Delta\varphi )$, which can be solved analytically. The discriminant of this polynomial 
vanishes at 
\begin{align}
    \left(\frac{B}{4C}\right)^2=\left(\frac{A+C}{3C}\right)^3,
    \label{discriminant}
\end{align}
indicating that two solutions disappear when $|B/C|$ exceeds some threshold value. In Appendix \ref{sec:appendix:parameter_discussion} we discuss such a case for a strongly asymmetric junction, in which case $|A|\ll |C|$ and the discriminant vanishes for $|B|\approx 4|C|/(3\sqrt{3})$ at a temperature of about 0.22 $T_c$. Near this temperature two $\widehat\Delta\varphi_\sn $ values approach each other, which allows to have a spin-diode efficiency close to 100\% in an extended range of $\Delta\varphi $. The value of $\Delta\varphi$ where Eq.~\eqref{discriminant} is met is given for $2\ge A/C\ge-1$ by
\begin{align}
    \widehat{\Delta\varphi}_\sn=\pm \arcsin \sqrt{\frac{A+C}{3C}},
    \label{discriminantvalue}
\end{align}
which for $|C|\gg |A|$ is approximately $\pm \arcsin(1/\sqrt{3})$.

To analyze the above analytical expressions for $\widehat\eta_{\cc,\sn}$, we perform extensive calculations in Appendix \ref{sec:appendix:parameter_discussion} for homogeneous superconducting order parameters with temperature-dependent BCS gap $\Delta(T)$ (as these calculations are numerically expensive, we do not determine the order parameter profile self-consistently). For each parameter combination we numerically calculate the spin-resolved CPR using the QCGF formalism and obtain the corresponding Fourier components which are used to calculate the spin and charge Josephson diode efficiency.
\begin{figure}[t]
    \centering
    \includegraphics[width=\linewidth]{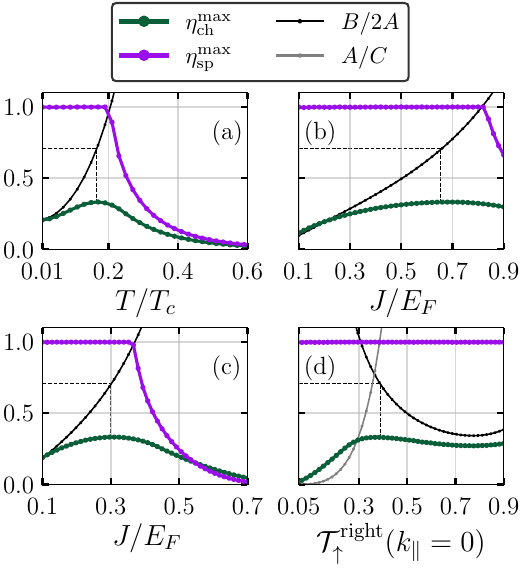}
    \caption{Functional dependence of maximum diode efficiencies $\eta^\mathrm{max}_{\cc,\sn}$ and the ratios $B/2A$ and $A/C$ [see Eqs.~\eqref{AB} and \eqref{eq:def_C}] on (a) the temperature,  (b) and (c) the exchange field strength for two different Fermi surface geometries [panel (b) has a Fermi surface mismatch as shown in Fig.~\ref{fig:sketch_parabolas}(b) and panel (c) as in Fig.~\ref{fig:sketch_parabolas}(c)], and (d) on the transmission probability of the right interface. The corresponding parameter sets are given in the respective Figs.~\ref{fig:appendix:symm:Diode_eff_var_temp}-\ref{fig:appendix:asymm:Diode_eff_var_transm} of Appendix~\ref{sec:appendix:parameter_discussion}.}
    \label{fig:B_2A_ratio_effs}
\end{figure}
In Fig.~\ref{fig:B_2A_ratio_effs} we show the maximum values $\eta_{\cc,\sn}^\mathrm{max}=\max_{\Delta\varphi} |\eta_{\cc,\sn}|$, calculated from the Fourier expansion by taking contributions up to $|\mu|+|\nu| \leq 2$, as function of (a) the temperature, (b)+(c) the exchange field strength, and (d) on the transmission probability of the right interface. In the following, we compare $\eta_{\cc,\sn}^\mathrm{max}$ with the analytical values $\widehat{\eta}_{\cc,\sn}$.
First we discuss the functional dependencies shown in panels (a)-(c) of Fig.~\ref{fig:B_2A_ratio_effs}. In agreement with the previous discussion the numerical results show that $\eta^\mathrm{max}_\sn = 1$ for $\abs{B} \leq 2\abs{A}$. For $\abs{B} > 2\abs{A}$ we obtain that $\eta^\mathrm{max}_\sn$ monotonically decreases as $B/2A$ increases. Also the numerical results for the maximal charge diode efficiency $\eta^\mathrm{max}_\cc$ verify the analytic calculations as the maximum magnitude of $\approx 33 \%$ is reached for $|B|=\sqrt{2}|A|$ (dashed lines in Fig.~\ref{fig:B_2A_ratio_effs}). We do not  show the $A/C$ ratio in panels (a)-(c) of Fig.~\ref{fig:B_2A_ratio_effs} because $|C|\ll |A|,|B|$ as the considered systems are symmetric. For panel (d) we vary the system from a symmetric junction $\mathcal{T}_\uparrow^\mathrm{right}(k_\parallel = 0)=0.9$ to a strongly asymmetric one $\mathcal{T}_\uparrow^\mathrm{right}(k_\parallel = 0)=0.05$. For a high transparency of the right FI layer we obtain $|B| < 2|A|$ and $|C| \ll |A|$ and thus Eq.~\eqref{eq:phi_sp} can be used. By decreasing the transparency of the right FI layer $A$ decreases whereas $C$ increases. Thus, for $0.3 \leq \mathcal{T}_\uparrow^\mathrm{right}(k_\parallel=0) \leq 0.4$ we obtain $|A| \sim |C|$ and $\eta_\cc^\mathrm{max}$ reaches its maximum value as $B/2A$ increases from below $1/\sqrt{2}$ to $B/2A > 1$. From panels (a)-(c) of Fig.~\ref{fig:B_2A_ratio_effs} we would expect that $\eta^\mathrm{max}_\sn < 1$ as $|B| > 2|A|$ but in panel (d) $\eta^\mathrm{max}_\sn = 1$ even for that case. This result follows from Eqs.~\eqref{eq:phi_sp} and \eqref{aroundpihalf} as for the strongly asymmetric case $|C| \gg |A|,|B|$ and therefore in the range $B>2A$ even two solutions with $|\widehat\eta_\sn|=1$ exist.
 
Comparing the results for the two different cases of Fermi surface mismatch shown in Figs.~\ref{fig:B_2A_ratio_effs}(b) and \ref{fig:B_2A_ratio_effs}(c), it is seen that for panel (b) the maximum charge diode efficiency is reached for $J \approx 0.65 E_F$ whereas in the other case the maximum is reached at $J \approx 0.3 E_F$. This effect of Fermi surface geometry leads us to an important observation. Scattering processes at the interface can be classified into type-(R): total reflection processes, type-(H): transmission only into one spin band of the sFM, and type-(F): transmission into both spin bands of the sFM. Scattering processes of type (H), which are of the type which occur at an interface with a half-metallic ferromagnet, appear in our model for $k_F^\downarrow < k_\parallel < \mathrm{min}(k_F^\uparrow,k_F)$. These processes only contribute to $I_{1,0}$, but not to $I_{0,1}$, thus increasing the magnitude of the quantity $I_{1-}=I_{1,0}-I_{0,1}$, which is essential for the Josephson diode effect. Consequently, for given exchange splitting of the electronic bands, the effects are maximized for a Fermi surface geometry of the type shown in Fig.~\ref{fig:sketch_parabolas}(c), for which $k_F^\uparrow \le k_F$. For a Fermi surface geometry of the type shown in  Fig.~\ref{fig:sketch_parabolas}(b), where $k_F^\uparrow \ge k_F$, only the part $k_F^\downarrow < k_\parallel <k_F$ supports type-(H) scattering processes, and the third type of Fermi surface geometry with $k_F^\downarrow >k_F$ (not shown in Fig.~\ref{fig:sketch_parabolas}) has no type-(H) scattering processes at all and therefore exhibits the least pronounced Josephson diode effects. We underline, however, that for Josephson junctions involving purely half-metallic ferromagnets, no Josephson diode effect occurs. Therefore, a trade-off between type-(H) and type-(F) scattering processes takes place, with the former giving important contributions to $I_{1-}$ and the latter to $I_{1,1}$ and $I_{1,-1}$.
The highest efficiencies occur when both types of processes contribute to about a similar degree.
 Thus, the difference between Figs.~\ref{fig:B_2A_ratio_effs}(b) and \ref{fig:B_2A_ratio_effs}(c) follows from the difference in exchange fields that is needed in order to have a comparable $k_\parallel$ range for which type-(H) scattering events take place. 
 
For a more detailed analysis of Josephson diode efficiencies as functions of temperature, exchange splitting between the bands, and interface transmission asymmetry, we refer to Appendix \ref{sec:appendix:parameter_discussion}.

Next, we discuss the quality of the model CPR in Eqs.~\eqref{eq:CPR_approx_upup}-\eqref{eq:CPR_approx_dodo} in comparison to the numerically obtained CPR's. 
\begin{figure}[th]
    \centering
    \includegraphics{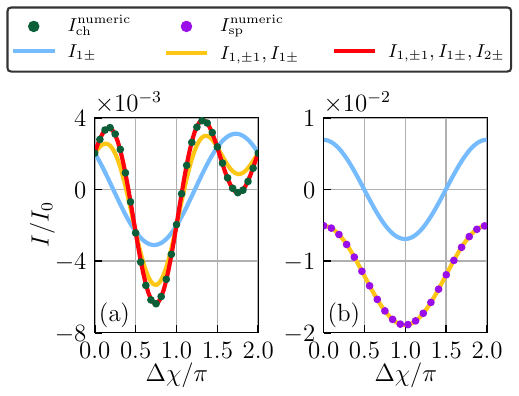}
    \caption{CPR and its approximations for (a) the charge current in a symmetric junction [see Fig.~\ref{fig:colormap_ch_sp_curr_and_spon_curr_diode_eff_SYMM}(a)] for $\Delta\varphi = 0.47 \pi$, and (b) for the spin current in an asymmetric junction [see Fig.~\ref{fig:colormap_sp_curr_diode_eff_ASYMM}(a)] for $\Delta\varphi = \pi/4$. The symbols show the currents numerically obtained from a self-consistent solution (only a few points are shown for a better visibility). The full lines are obtained from the approximations in Eqs.~\eqref{Icappr}-\eqref{Isappr}. Different colors take into account different numbers of Fourier contributions as denoted in the legend.}
    \label{fig:CPR_approx}
\end{figure}
From Fig.~\ref{fig:CPR_approx}(a) it follows that taking into account all contributions shown in Eq.~\eqref{Icappr} gives a quantitative description of the charge current. Considering only the contributions for $\abs{\mu},\abs{\nu} \leq 1$ is not sufficient [see Fig.~\ref{fig:CPR_approx}(a)] since the $I_{2,0}$ and $I_{0,2}$ contributions are of a comparable order of magnitude [see Figs.~\ref{fig:Fourier_comp}(a),(e), and (g)]. In contrast, we obtain for the spin current, shown in Fig.~\ref{fig:CPR_approx}(b), that for its approximation it is sufficient to account for the contributions where at maximum one pair is transmitted through each spin-channel. We obtain that the functional dependence on $\Delta\chi$ is dominated by the $I_{1+}$ contribution, i.e., $\cos(\Delta\chi)$. The $I_{1-}$ contribution is comparatively small, but the $I_{1,-1}$ contribution is comparable to the $I_{1+}$ contribution and yields a $\Delta\varphi$-dependent shift of the CPR.

Finally, we remark that the processes for which an equal number of pairs is transmitted through each spin channel, i.e., $\mu = \nu$, are the \textit{crossed pair transmission} contributions \cite{greinSpinDependentCooperPair2009}. The leading term, $I_{1,1}$, dominates the higher harmonics which are crucial for the Josephson diode effect. This term describes a simultaneous transmission of one equal-spin pair in each spin band. 
Furthermore, Eq.~\eqref{Isappr} shows that the spin CPR is a superposition between a term constant in $\Delta \chi$ and various oscillating terms. In the case that the magnitude of $2I_{1,-1}\sin(2\Delta\varphi)$ becomes larger than the amplitudes of the oscillating terms (which happens in strongly asymmetric junctions), the spin current flows only in one direction, irrespective of the superconducting phase difference $\Delta \chi$. Were one to define a spin diode efficiency via the maximum and minimum spin currents, then a 100\% efficiency would be present for entire ranges of $\Delta\varphi$ in such junctions.
However, such a spin current even flows when the Josephson effect is absent, i.e., when one of the two interfaces is intransparent. In this case the charge current is identically zero, such that only terms with $\mu+\nu=0$ contribute to the Fourier sums in Eqs.~\eqref{eq:I_uu}-\eqref{eq:I_dd}. For such a junction, $I_\sn= -\sum_{\mu=0}^\infty 2\mu I_{\mu,-\mu} \sin (2\mu \Delta \varphi)$ \cite{eschrigSpinpolarizedSupercurrentsSpintronics2015,eschrigPhasesensitiveInterfaceProximity2019}.  Therefore, we define the spin Josephson diode effect via the values of the spin current at the phases corresponding to the critical charge currents. In this case, the spin diode efficiency is directly related to the Josephson effect. As we have shown in this article, spin diode efficiency of 100\% can still be reached with this definition, namely when for a certain value of $\Delta\varphi $ the spin current vanishes for either $\Delta \chi_\cc^+$ or $\Delta\chi_\cc^-$.

\subsection{Spontaneous currents}
The system carries a spontaneous current even in the absence of a superconducting phase difference at the outer interfaces of the structure. This situation corresponds to a loop geometry with an inserted SC/sFM/SC junction, where the spontaneous current gives rise to a spontaneous flux of $\frac{\Delta\chi_0}{2\pi}\Phi_0$ through the loop \cite{bauerSpontaneousSupercurrentInduced2004}. Here, $\Delta\chi_0$ is the phase difference between the two SC/sFM interfaces and $\Phi_0$ is the superconducting flux quantum. The former depends on $\Delta\varphi$ and appears only if $\Delta\varphi \ne k\pi$, $k\in\mathbb{Z}$.

\begin{figure}[b]
    \centering
    \includegraphics[width=\linewidth]{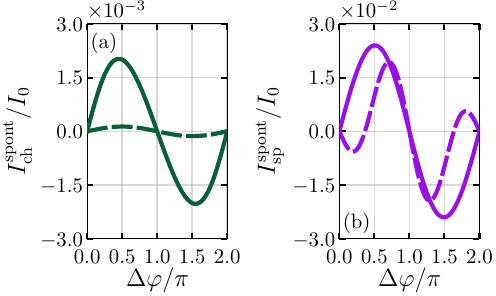}
    \caption{The functional dependence of (a) the spontaneous charge current and (b) the spontaneous spin current on the geometric phase difference $\Delta\varphi$ for the symmetric system considered in Fig.~\ref{fig:colormap_ch_sp_curr_and_spon_curr_diode_eff_SYMM} (solid lines) and the asymmetric system from Fig.~\ref{fig:colormap_sp_curr_diode_eff_ASYMM} (dashed lines), respectively.}
    \label{fig:combined_spon_CPR}
\end{figure}
\begin{figure*}[t]
    \centering
    \includegraphics[width=\linewidth]{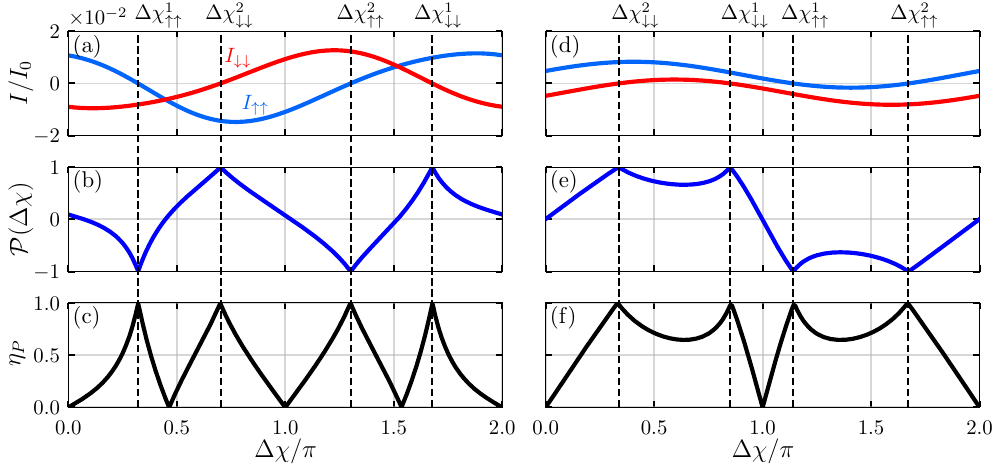}
    \caption{The functional dependence of the spin-polarization $\mathcal{P}$ [see Eq.~(\ref{eq:curr_spin_polarization})] on the superconducting phase difference $\Delta\chi$ for a symmetric junction (b) and an asymmetric junction (e) for the spin-resolved currents shown in (a) and (d), respectively. In (c) and (f) the corresponding switching efficiencies $\eta_\mathcal{P}$ are shown [see Eq.(\ref{eq:switching_efficiency})]. For the symmetric case (left column) $\Delta\varphi=0.31\pi$ and for the asymmetric case (right column) $\Delta\varphi=0.91\pi$.}
    \label{fig:CURR_POL}
\end{figure*}
The numerically obtained self-consistent spontaneous charge (green) and spin (violet) currents are shown in Fig.~\ref{fig:combined_spon_CPR}. 
For the symmetric case, both the spontaneous spin and charge current are approximately sinusoidal [see solid lines in Fig.~\ref{fig:combined_spon_CPR}]. In contrast, the spontaneous spin current in an asymmetric junction shows a strong second harmonic contribution whereas the charge current is strongly suppressed. This is due to the smaller transparency of the right interface. 

We can understand these results in terms of the analytical model introduced in Sec. \ref{sec:analytic_model}.
The spontaneous current is obtained from Eqs.~\eqref{Icappr}-\eqref{Isappr} by setting $\Delta\chi=0$. Neglecting the small terms proportional to $I_{2\pm}$, this leads to
\begin{align}
    I_\cc^{\rm spont} \approx &
 I_{1-}\sin\qty(\Delta\varphi)
    \label{Icspon} \\
    I_\sn^{\rm spont} \approx &-2 I_{1,-1} \sin\qty(2\Delta\varphi)
     +I_{1+}\sin\qty(\Delta\varphi).
     \label{Isspon}
\end{align}
From this it is expected that the spontaneous charge current is approximately sinusoidal in $\Delta\varphi$, whereas the spontaneous spin current is approximately sinusoidal in symmetric junctions, where $I_{1,-1}\ll I_{1+}$, however has pronounced second harmonics in strongly asymmetric junctions, where $I_{1,-1}\sim I_{1+}$. Both spontaneous currents require that $\Delta\varphi $ is not a multiple of $\pi$, i.e., they require a {\it noncoplanar} spin arrangement.
The magnitude of the spontaneous charge current is given approximately by $I_{1-}=I_{1,0}-I_{0,1}$, and therefore  its presence requires different densities of states in the two spin bands. The magnitude of the spontaneous spin current, on the other hand, is of the order of $I_{1+}=I_{1,0}+I_{0,1}$, and only requires noncoplanar spin arrangement. The strong second harmonic in $\Delta\varphi$ proportional to $I_{1,-1}$ requires two spin-bands to be present - in a structure with a half-metallic ferromagnet this term is absent.

\subsection{Spin-polarization switch}
Finally, we discuss an effect present in both junction types which we consider of being of interest for experimental investigations. Typically, the geometric phase difference $\Delta\varphi$ is fixed during operation of the junction. Therefore, we now consider the systems for fixed $\Delta\varphi$ but varying $\Delta\chi$ which may be realized by embedding the ferromagnetic trilayer considered throughout in a SQUID-geometry and applying a magnetic flux $\Phi$ through the loop. Considering the spin-resolved currents for different $\Delta\varphi$'s in Figs.~\ref{fig:CPR_diode_effect_symm}(b), \ref{fig:CURR_POL}(a), and \ref{fig:CURR_POL}(d), it becomes evident that there is a value of the superconducting phase difference $\Delta\chi_{\uparrow\uparrow} (\Delta\varphi)$ for which $I_\upup(\Delta\chi_{\upup}) = 0$ but $I_\dodo(\Delta\chi_\upup) \neq 0$. Analogously, there is a value $\Delta\chi_\dodo (\Delta\varphi)$ for which $I_\dodo(\Delta\chi_{\dodo}) = 0$ but $I_\upup(\Delta\chi_\dodo) \neq 0$.

For a quantitative discussion, we define the spin polarization of the current as follows:
\begin{equation}
    \mathcal{P}(\Delta\chi) = \frac{\abs{I_\upup(\Delta\chi)} - \abs{I_\dodo(\Delta\chi)}}{\abs{I_\upup(\Delta\chi)} + \abs{I_\dodo(\Delta\chi)}}. \label{eq:curr_spin_polarization}
\end{equation}
Apparently, for $\Delta\chi = \Delta\chi_{\upup} $ and $\Delta\chi = \Delta\chi_{\dodo} $ it takes values $1$ and $-1$, respectively, corresponding to a fully spin-polarized current. 

As an example let us consider first the spin-resolved currents in a symmetric junction shown in Fig.~\ref{fig:CURR_POL}(a). The spin polarization of the Josephson current [see Eq.~\eqref{eq:curr_spin_polarization}] for such a situation is presented in panel (b) which displays the characteristic peaks at $\Delta\chi_{\upup (\dodo)}^{(1,2)}$ as expected from the definition and the index $1,2$ denotes the considered branch.

To get a qualitative insight into the switching effects, let us consider the approximate formulas given in the preceding subsection [see Eqs.~\eqref{eq:CPR_approx_upup} and \eqref{eq:CPR_approx_dodo}] assuming that $\abs{I_{1,-1}}, \abs{I_{2,0}}, \abs{I_{0,2}}< \abs{I_{1,1}} < \abs{I_{0,1}},\abs{I_{1,0}}$ which is usually the case for symmetric junctions [see Fig.~\ref{fig:Fourier_comp}]. Thus, the dominating terms for symmetric junctions are of the form 
\begin{align}
    I_\upup &\approx -I_{1,0}\sin(\Delta\chi-\Delta\varphi) \\I_\dodo &\approx -I_{0,1}\sin(\Delta\chi+\Delta\varphi).
\end{align}
 As a result, and as can be seen in Fig.~\ref{fig:CURR_POL}(a),
the spin-resolved currents for symmetric junctions diminish at $\Delta\chi_{\upup(\dodo)}^{(1)}\approx \pm \Delta\varphi$ and ${\Delta\chi_{\upup(\dodo)}^{(2)}\approx \pm \Delta\varphi + \pi}$. 
Furthermore, we obtain that the polarization of the current can be switched by changing the superconducting phase difference from $\Delta\chi_\upup^{(1)}$ to $\Delta\chi_\dodo^{(1)} \approx -\Delta\chi_\upup^{(1)}$. In a SQUID geometry this transition is equivalent to a reversal of the flux direction, i.e., $\Phi \rightarrow - \Phi$. This effect is interesting for future experimental investigations and may lead potentially to new types of devices.
 To numerically quantify the switching effect, we introduce the following quantity: 
\begin{equation}
    \eta_\mathcal{P}(\Delta\chi) = \frac{\abs{P(\Delta\chi)-P(-\Delta\chi)}}{2}, \label{eq:switching_efficiency}
\end{equation}
which we term the {\it switching efficiency}. The functional dependence of $\eta_\mathcal{P}(\Delta\chi)$ on the superconducting phase difference $\Delta\chi$ resulting from the CPR in Fig.~\ref{fig:CURR_POL}(a) is shown in Fig.~\ref{fig:CURR_POL}(c). In contrast to the polarization in Fig.~\ref{fig:CURR_POL}(b), the switching efficiency peaks at $\pm (\Delta\chi^{(1,2)}_{\upup} - \Delta\chi^{(1,2)}_{\dodo})/2$, i.e., slightly besides $\Delta\chi^{(1,2)}_{\upup}$ and $\Delta\chi^{(1,2)}_{\dodo}$, and reaches up to $99 \%$.

For asymmetric junctions, on the other hand, we have rather $\abs{I_{2,0}}, \abs{I_{0,2}}, \abs{I_{1,1}} < \abs{I_{1,0}}, \abs{I_{0,1}}, \abs{I_{1,-1}}$, and therefore the dominating terms in Eqs.~\eqref{eq:CPR_approx_upup} and \eqref{eq:CPR_approx_dodo} are 
\begin{align}
    I_\upup &\approx -I_{1,0}\sin(\Delta\chi-\Delta\varphi)-I_{1,-1}\sin(2\Delta\varphi)\\
    I_\dodo &\approx -I_{0,1}\sin(\Delta\chi+\Delta\varphi)+I_{1,-1}\sin(2\Delta\varphi).
\end{align}
Thus, both have an additional constant term (independent of $\Delta\chi$), which for $\Delta\varphi \ne k\pi/2$ shifts the CPRs in vertical direction, modifying the zero crossings.
As a consequence, we obtain a switching effect for asymmetric junctions as shown in  Figs.~\ref{fig:CURR_POL}(e) and \ref{fig:CURR_POL}(f); although qualitatively similar to the case of symmetric junctions, quantitatively the functional dependencies are different and three cases can be distinguished. Firstly, for certain $\Delta\varphi$ we obtain the same effect as discussed for the symmetric junction, i.e., $\mathcal{P}$ and $\eta_\mathcal{P}$ peak at $\Delta\chi^{(1/2)}_{\upup(\dodo)}$ and reach zero  between two peaks. Secondly, if $\Delta\varphi$ is chosen such that neither $I_\upup$ nor $I_\dodo$ change their sign as function of $\Delta\chi$, $\mathcal{P}$ does not reach $\pm1$ and consequently $\eta_\mathcal{P} < 1$ for all $\Delta\chi$. Finally, when $I_\upup$ and $I_\dodo$ do not intersect, i.e., $I_\upup \ne I_\dodo$ for all $\Delta\chi$ [see Fig.~\ref{fig:CURR_POL}(d)], we obtain that $\mathcal{P} = \pm 1$ and $\eta_\mathcal{P} \approx 99\%$ at $\Delta\chi^{(1/2)}_{\upup(\dodo)}$ but the polarization and the switching efficiency only vanish for $\Delta\chi = k\pi$, $k\in\mathbb{Z}$. Thus, for $\Delta\chi \in [\Delta\chi_{\uparrow\uparrow}^{1},\Delta\chi_{\uparrow\uparrow}^{2}]$ and $\Delta\chi \in [\Delta\chi_{\downarrow\downarrow}^{2},\Delta\chi_{\downarrow\downarrow}^{1}]$ the polarization $\mathcal{P}$ and $\eta_\mathcal{P}$ remain at comparatively high values. Consequently, the superconducting phase difference, and thus the flux $\Phi$, do not need to be adjusted too accurately to $\Delta\chi^{(1/2)}_{\upup(\dodo)}$ to achieve a highly spin-polarized current and a high switching efficiency. Hence, even though the currents are smaller in the asymmetric junction, it might be easier to exploit this effect in applications.

\section{Conclusion}\label{sec:Conclusion}
In summary, we have presented a systematic theoretical analysis of the Josephson diode effect in macroscopic hybrid junctions comprising a strongly spin-polarized ferromagnet sandwiched between two BCS superconducting leads via two spin-active insulating interfaces. The study has been delivered in the framework of modified quasiclassical Green's function formalism in the diffusive limit adapted to strongly spin-polarized materials. The strong exchange field in the sFM allows to neglect the mixed-spin correlations (spin singlets and spin triplets with $s_z=0$) yielding only equal-spin triplets ($s_z=\pm 1$). This regime allows us to get an insight in spin-resolved quantities and accordingly calculate the transport properties.  

We have focussed on the role of the relative azimuthal angle between the magnetizations of the two ferromagnetic interfaces, which plays the role of a \mbox{\textit{quantum-geometric phase difference}}, $\Delta\varphi$, by entering directly the Josephson current phase relation. We have shown that the Josephson diode effect can be realized if the following conditions are fulfilled: (i) noncoplanar spin texture, i.e., nonvanishing $\Delta\varphi$ (but $\Delta\varphi \neq k\pi/2$) and (ii) different densities of states between the spin bands (i.e., Fermi surface spin splitting). Both spin bands are required in order to enable prominent crossed pair transmission processes, i.e., the effects are absent in half-metals. The mentioned requirements are fulfilled in the case of a strongly spin-polarized ferromagnet and a noncoplanar profile of magnetizations in the trilayer. A strong spin polarization of the central region destroys any  coherence of pair correlations that involve both spin bands, allowing only equal-spin pairs amplitudes to enter either spin-$\uparrow$ or spin-$\downarrow$ bands. Our model predicts charge diode efficiency of up to $\eta_\cc \gtrsim 33\%$ and a perfect spin diode effect with $\eta_\sn=100\%$.

To get more insight into the role of the geometric phases, we have performed the harmonic analysis of the spin-resolved currents both in $\Delta\chi$ and $\Delta\varphi$, where the former denotes the superconducting phase difference. We discussed symmetries between the Fourier coefficients and derived a physically appealing model in which $\Delta\chi$ and $\Delta\varphi$ enter the theory on equal footing. 

Beside the diode effects, we have analyzed the spin switching effects in the system. We have shown that the SQUID geometry involving a ferromagnetic trilayer can feature fully spin-polarized supercurrents. In addition, we have shown that its polarization can be well controlled by an external magnetic flux. 

Finally, our model can serve as a platform for further experimental investigations of the Josephson charge and spin diode effect in mesoscopic hybrids involving strongly spin-polarized metallic ferromagnets based on Ni or Co~\cite{khaireObservationSpinTripletSuperconductivity2010,Glick2018,Aguilar2020}, and ferromagnetic insulators based on GdN \cite{senapatiSpinfilterJosephsonJunctions2011,Caruso2019} or EuS/EuO~\cite{mooderaPhenomenaSpinfilterTunnelling2007,yangPulsedLaserDeposition2014,dieschCreationEqualspinTriplet2018}. In addition, the discussed system can be a promising platform for experimental investigation of spin switching effects in SQUID devices involving ferromagnetic trilayers.

\acknowledgements
We thank Ralf Schneider for fruitful discussions. NLS and ME acknowledge funding by the Deutsche Forschungsgemeinschaft (DFG, German Research
Foundation) under project number 530670387. The computations were enabled by resources provided by the University Computer Centre of the University of Greifswald.
\section*{Data Availablitiy}
The data that support the findings of this article are openly available \cite{data}.
\appendix

\section{Calculation of the Fourier coefficients} \label{sec:appendix:Fourier}
As shown in the the main text, noncoplanar spin textures in SC/sFM/SC trilayers introduce the quantum-geometric phase difference $\Delta\varphi$, which enters the current-phase relations. Therefore, a Fourier decomposition of the spin-resolved currents both in the superconducting phase difference $\Delta\chi=\chi_2-\chi_1$ and the geometric phase difference $\Delta\varphi=\varphi_2-\varphi_1$  can be performed to gain more insights. 

Starting from the Fourier expansion Eq.~\eqref{eq:general_Fourier_ansatz} we obtain
\begin{equation}
\label{eqn:Fourier_AB}
\begin{split}
    I_{\eta\eta}(\Delta\chi,\Delta\varphi) = \frac{1}{4} \sum_{m,n=-\infty}^\infty &A_{m,n}^\eta \sin(m\Delta\chi) \cos(n\Delta\varphi) \\
    + &B_{m,n}^\eta \cos(m\Delta\chi) \sin(n\Delta\varphi),
\end{split}
\end{equation}
where we formally extended the Fourier sums over $n$ and $m$ to $-\infty $ using the definitions 
\begin{align}
A^\eta_{-m,n} &= -A^\eta_{m,n}, & \quad A^\eta_{m,-n}&= A^\eta_{m,n},
\label{Asymm}
\\ B^\eta_{-m,n}&= B^\eta_{m,n}, & \quad B^\eta_{m,-n} &= -B^\eta_{m,n} .
\label{Bsymm}
\end{align}
In particular, it follows that $A^\eta_{0,n}=0$, $B^\eta_{m,0}=0$. Eq.~\eqref{eqn:Fourier_AB} can be rewritten using the trigonometric identity $\sin \alpha \cos \beta = \frac{1}{2}\left[ \sin (\alpha+\beta)+\sin (\alpha-\beta)\right]$,  as 
\begin{align}
    I_{\eta\eta} 
    = \frac{1}{4} \sum_{m,n=-\infty}^\infty &\qty(A_{m,n}^\eta + B_{m,n}^\eta) \sin(m\Delta\chi + n \Delta\varphi) \nonumber \\
    \quad+ &(A_{m,n}^\eta-B_{m,n}^\eta) \sin(m\Delta\chi-n\Delta\varphi).
\end{align}
Renaming the summation variable $n$ in th first line into $-n$ and using Eqs.~\eqref{Asymm}-\eqref{Bsymm}
leads with the definition $\mathcal{J}_{m,n}^{\eta} = (A_{m,n}^\eta - B_{m,n}^\eta)/2$ to
\begin{equation}
    I_{\eta\eta}(\Delta\chi,\Delta\varphi) = \frac{1}{2}\sum_{m,n=-\infty}^\infty \mathcal{J}_{m,n}^{\eta}\sin(m\Delta\chi - n \Delta\varphi), \label{eq:fourier:symmetry_constrained}
\end{equation}
where $\mathcal{J}^{\eta}_{m,n} = -\mathcal{J}^{\eta}_{-m,-n}$.
The numerical current-phase relations allow for the calculation of the Fourier components $A_{m,n}^\eta$ and $B_{m,n}^\eta$ which are shown for the case of a symmetric and a asymmetric junction configuration in Fig.~\ref{fig:Fourier_comp}. Explicit numerical evaluation confirms that the Fourier components are restrained by additional symmetry relations, some of which are summarized in Table~\ref{tab:FT_symmetries}, and which can be summarized by the relation
\begin{equation}
    n\left(\mathcal{J}_{m,n}^{\uparrow} + \mathcal{J}_{m,n}^{\downarrow}\right)=m\left(\mathcal{J}_{m,n}^{\uparrow} - \mathcal{J}_{m,n}^{\downarrow}\right).
\end{equation}
Our analysis leads to nonzero Fourier coefficients only for combinations of $m$ and $n$ where $m\pm n$ are even integers. We therefore pass to 
new integer indices $(\mu,\nu)$ defined as $\mu=(m+n)/2$ and $\nu=(m-n)/2$. Accordingly, we define $I_{\mu,\nu}^{\eta} \equiv  \mathcal{J}_{\mu + \nu,\mu - \nu}^{\eta}$. 
Consequently, the corresponding symmetry relations are shown in the right column of Table~\ref{tab:FT_symmetries} and the above relation for $\mathcal{J}_{m,n}^\eta$ reads 
\begin{equation}
(\mu-\nu)\left(I_{\mu,\nu}^{\uparrow} + I_{\mu,\nu}^{\downarrow}\right)=(\mu+\nu) \left(I_{\mu,\nu}^{\uparrow} - I_{\mu,\nu}^{\downarrow}\right), \label{eq:fraction_relation_I_ij}
\end{equation}
or, equivalently,
\begin{equation}
\mu  I_{\mu,\nu}^{\downarrow}= \nu I_{\mu,\nu}^{\uparrow}. \label{eq:fraction_relation_I_ij_2}
\end{equation}
Defining $I_{\mu,\nu}^{\uparrow} = \mu  \tilde I_{\mu,\nu}$, this leads to
\begin{equation}
I_{\mu,\nu}^{\uparrow} = \mu\cdot \tilde I_{\mu,\nu} \quad \mathrm{and} \quad I_{\mu,\nu}^{\downarrow} = \nu\cdot \tilde I_{\mu,\nu},
\end{equation}
where $\tilde I_{-\mu,-\nu} = \tilde I_{\mu,\nu}$. This allows to rewrite the initial Fourier ansatz, see Eq.~(\ref{eqn:Fourier_AB}), as 
\begin{align}
\begin{split}
    I_{\uparrow\uparrow} &= \frac{1}{2}\sum_{\mu,\nu=-\infty}^\infty \mu \tilde I_{\mu,\nu} \sin\qty[(\mu+\nu)\Delta\chi - (\mu-\nu)\Delta\varphi], \\
    I_{\downarrow\downarrow} &= \frac{1}{2}\sum_{\mu,\nu=-\infty}^\infty \nu \tilde I_{\mu,\nu} \sin\qty[(\mu+\nu)\Delta\chi - (\mu-\nu)\Delta\varphi],
\end{split}
\label{IupupIdowndown}
\end{align}
which are Eqs.~\eqref{eq:I_uu} and \eqref{eq:I_dd} from the main text, where we wrote $\tilde I_{\mu,\nu}=(-1)^{\mu+\nu}I_{\mu,\nu}$.
\begin{table}[t!]
\label{tab:FT_symmetries}
\caption{Tabulated symmetries between $A^\eta_{m,n}$ and $B^\eta_{m,n}$ coefficients [see Eq.~\eqref{eqn:Fourier_AB}], for the $\mathcal{J}_{m,n}^{\eta}$ coefficients [see Eq.~\eqref{eq:fourier:symmetry_constrained}], and the  $I_{\mu,\nu}^\eta$'s.} 
    \begin{tabular}{|c|c|c|}
    \hline
    \multicolumn{3}{|c|}{Symmetries in} \\
    \hline
    \text{$A_{m,n}^\eta$ and $B_{m,n}^\eta$} & \text{$\mathcal{J}_{m,n}^{\eta}$} &
    \text{$I_{\mu,\nu}^{\eta}$} \\
    \hline\hline
    $A_{m,0}^\uparrow = A_{m,0}^\downarrow$ & $\mathcal{J}_{m,0}^{\uparrow} = \mathcal{J}_{m,0}^{\downarrow}$ & $I_{\mu,\mu}^{\uparrow} = I_{\mu,\mu}^{\downarrow}$ \\
    \hline
    $B_{0,n}^\uparrow = -B_{0,n}^\downarrow$  & $\mathcal{J}_{0,n}^{\uparrow} = -\mathcal{J}_{0,n}^{\downarrow}$ &
    $I_{\mu,-\mu}^{\uparrow} = - I_{\mu,-\mu}^{\downarrow}$ \\
    \hline
    $A_{m,m}^\uparrow = -B_{m,m}^\uparrow$ & $\mathcal{J}_{m,-m}^{\uparrow} = 0$ &
    $I_{0,\nu}^{\uparrow} = 0$\\
    \hline
    $A_{m,m}^\downarrow = B_{m,m}^\downarrow$ & $\mathcal{J}_{m,m}^{\downarrow} = 0$ & $I_{\mu,0}^{\downarrow} = 0$\\
    \hline
    \end{tabular}
\end{table}

\section{Boundary conditions for the coherence amplitudes} \label{sec:appendix:boundaries}
The normalization of the quasiclassical Green's function allows to express it in terms of projectors $\hat{P}_\pm$~\cite{shelankovDerivationQuasiclassicalEquations1985,eschrigDistributionFunctionsNonequilibrium2000}
\begin{equation}
    \hat{G} = (\hat{P}_+ - \hat{P}_-),
\end{equation}
where $\hat{P}_\pm$ is defined as
\begin{equation}
    \hat{P}_\pm = \frac{1}{2}\left(\hat{1} \pm \hat{G}\right).
\end{equation}
It can be easily shown that these are projectors having the following properties:
\begin{subequations}
\begin{align}
 \hat{P}_\pm^2 &= \hat{P}_\pm, \label{Id3a}\\
    \hat{P}_+ \hat{P}_- &= \hat{P}_- \hat{P}_+=\hat{0}, \label{Id3b}\\
    \hat{P}_+ + \hat{P}_- &= \hat{1},
\end{align}
\end{subequations}
implying
\begin{align}
    \pdv{\hat{P}_+}{z} &= - \pdv{\hat{P}_-}{z}, \\
    \pdv{\hat{P}_\pm}{z}\hat{P}_\mp &= -\hat{P}_\pm \pdv{\hat{P}_\mp}{z}, \\
    \hat{P}_\pm \pdv{\hat{P}_\mp}{z} \hat{P}_\pm &= \hat{0}.
\end{align}
Using the above expression, the following identity can be easily derived
\begin{align}
    \pdv{\hat{P}_+}{z} &= \qty(\hat{P}_+ + \hat{P}_-) \pdv{\hat{P}_+}{z} \qty(\hat{P}_+ +\hat{P}_-) \nonumber\\
    &= \hat{P}_+ \pdv{\hat{P}_+}{z} \hat{P}_- + \hat{P}_- \pdv{\hat{P}_+}{z} \hat{P}_+. \label{Id1}
\end{align}
An analogous expression can be derived for $\hat{P}_-$ by exchanging $+$ and $-$ indices in the above equation.

The boundary condition for the Green's function Eq.~(\ref{eq:final_boundary_cond_GF}) can be written in terms of projectors as 
\begin{equation}
   \pdv{\hat{P}_+}{z} - \pdv{\hat{P}_-}{z} = 2\pdv{\hat{P}_+}{z} = -\frac{1}{r}\qty(\hat{P}_+ - \hat{P}_-) \hat{\overline{\mathcal{I}}} ,
\end{equation}
which by myltiplying from the left with $\hat{P}_+$ and using Eqs.~\eqref{Id1}, \eqref{Id3a} and \eqref{Id3b} is rewritten as 
\begin{align}
     \hat{P}_+ \pdv{\hat{P}_+}{z} \hat{P}_- &= -\frac{1}{2r}\hat{P}_+ \hat{\overline{\mathcal{I}}}. \label{eq:P+}
\end{align}
An analogous calculation leads to 
\begin{equation}
     \hat{P}_- \pdv{\hat{P}_-}{z} \hat{P}_+ = -\frac{1}{2r} \hat{P}_-  \hat{\overline{\mathcal{I}}}.\label{eq:P-}
\end{equation}
Multiplying Eq.~\eqref{eq:P+} from the right with $\hat{P}_+$ and Eq.~\eqref{eq:P-} from the right with
$\hat{P}_-$ yields 
\begin{equation}
    \hat{P}_\pm \hat{\overline{\mathcal{I}}} \hat{P}_\pm = \hat{0}. \label{eq:symm_projectors}
\end{equation}

Since we use coherence (Riccati) amplitudes in this work, we express the above relation in its terms. We start from projectors that read as follows \cite{eschrigDistributionFunctionsNonequilibrium2000}:
\begin{align}
    \hat{P}_+ &= \begin{pmatrix} \mathit{1} \\ -\tilde{\gamma} \end{pmatrix}
    \; N\; 
    (\mathit{1},\gamma), \\
    \hat{P}_- &= \begin{pmatrix} -\gamma \\ \mathit{1} \end{pmatrix}
    \; \tilde N \;
    (\tilde{\gamma},\mathit{1}), \\
    N&=(\mathit{1}-\gamma\tilde{\gamma})^{-1},\quad \tilde N=(\mathit{1}-\tilde{\gamma}\gamma)^{-1}.
\end{align}
Eq.~\eqref{eq:symm_projectors} then takes the form 
\begin{align}
\overline{\mathcal{I}}_{11}+\gamma \overline{\mathcal{I}}_{21}-\overline{\mathcal{I}}_{12}\tilde\gamma -\gamma \overline{\mathcal{I}}_{22}\tilde\gamma&=0,\\
\overline{\mathcal{I}}_{22}+\tilde\gamma \overline{\mathcal{I}}_{12}-\overline{\mathcal{I}}_{21}\gamma
 -\tilde\gamma \overline{\mathcal{I}}_{11} \gamma&=0 .
\end{align}
The left hand side of Eq.~\eqref{eq:P+} adopts the form
\begin{equation}
    \hat{P}_+ \pdv{\hat{P}_+}{z} \hat{P}_- = \begin{pmatrix} \mathit{1} \\ -\tilde{\gamma} \end{pmatrix} 
    \; N\; \pdv{\gamma}{z} \ 
    \; \tilde N \; (\tilde{\gamma},\mathit{1}).
\end{equation}
Similarly the $\hat{P}_+\hat{\mathcal{I}}$ term on the right hand side of the same equation can be evaluated, finally, leading to the boundary conditions
\begin{widetext}
\begin{equation}
    \begin{pmatrix} N \pdv{\gamma}{z} \tilde{N} \tilde{\gamma} & N \pdv{\gamma}{z} \tilde{N} \\ -\tilde{\gamma} N \pdv{\gamma}{z} \tilde{N} \tilde{\gamma} & - \tilde{\gamma} N \pdv{\gamma}{z} \tilde{N} \end{pmatrix} = -\frac{1}{2r} \begin{pmatrix} N \overline{\mathcal{I}}_{11} + N\gamma\overline{\mathcal{I}}_{21} & N\overline{\mathcal{I}}_{12} + N\gamma\overline{\mathcal{I}}_{22} \\ -\tilde{\gamma}N\overline{\mathcal{I}}_{11} - \tilde{\gamma} N \gamma \overline{\mathcal{I}}_{21} & -\tilde{\gamma}N\overline{\mathcal{I}}_{12} - \tilde{\gamma} N \gamma \overline{\mathcal{I}}_{22} \end{pmatrix},
\end{equation}
\end{widetext}
Note that in the above equation the various matrix components are not independent, and only one condition follows from them.
Since we are free to choose one equation, we take the (12)-component of the above equation and rewrite it as follows:
\begin{align}
    -2r\pdv{\gamma}{z} &= \qty(\overline{\mathcal{I}}_{12} + \gamma\overline{\mathcal{I}}_{22})\tilde{N}^{-1}.
\end{align}
A boundary condition for $\partial\tilde{\gamma}/\partial z$ is obtained by considering $\hat{P}_-$ instead of $\hat{P}_+$ in Eq.~(\ref{eq:boundary_projector}) which results in
\begin{equation}
    -2r\pdv{\tilde{\gamma}}{z} = \qty(\overline{\mathcal{I}}_{21} + \tilde{\gamma}\overline{\mathcal{I}}_{11}) N^{-1}.
\end{equation}
By utilizing Eq.~\eqref{eq:symm_projectors}, the above boundary conditions can be rewritten as 
\begin{align}
    -2r \pdv{\gamma}{z} &= N^{-1} \qty(\overline{\mathcal{I}}_{12} - \overline{\mathcal{I}}_{11} \gamma), \\
    -2r \pdv{\tilde{\gamma}}{z} &= \tilde{N}^{-1} \qty(\overline{\mathcal{I}}_{21} - \overline{\mathcal{I}}_{22}\tilde{\gamma}).
\end{align}

\section{Functional dependence of diode efficiencies and spontaneous currents on the system parameters}
\label{sec:appendix:parameter_discussion}
In the main text we have concentrated on discussing the Josephson diode efficiency on the geometric phase difference $\Delta\varphi$. 
In this Appendix we discuss the Josephson diode efficiencies as function of additional system parameters such as temperature, exchange splitting of the electronic bands in the ferromagnet, or transmission probability asymmetry between the left and right interfaces of the ferromagnet. As the numerical calculations are very extensive, we
consider for the following numerical results homogeneous superconductors (no spatial variation of the order parameter) to each side of the strongly spin-polarized ferromagnet. The homogeneous superconductors are characterized by a temperature dependent BCS gap $\Delta(T)$. We first discuss charge and spin diode efficiencies and proceed afterwards to spontaneous currents in a loop geometry.

\begin{figure}[t!]
    \centering
    \includegraphics[width=\linewidth]{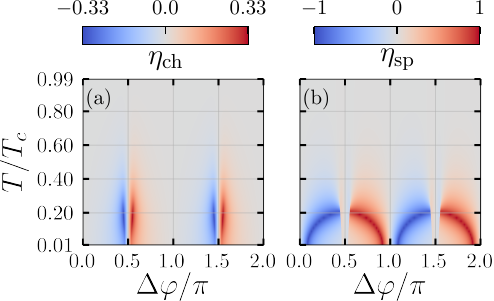}
    \caption{The functional dependence of the charge [panel (a)] and spin [panel (b)] diode efficiencies $\eta_{\cc/\sn}$ on the geometric phase $\Delta\varphi$ and the temperature $T$ of a symmetric Josephson junction. Note that the superconducting order parameter $\Delta(T)$ also varies with temperature. The remaining parameters are as in Fig. \ref{fig:CPR_diode_effect_symm} except that the ferromagnetic insulating layer on the left and right side have now the same width for both spin-channels which is $0.6\lambda_F/2\pi$. As in the main text the diode efficiencies are calculated by using the Fourier expansion of the currents [see Eqs.~\eqref{eq:I_uu} and \eqref{eq:I_dd}] for which the numerically obtained Fourier contributions were utilised.}
    \label{fig:appendix:symm:Diode_eff_var_temp}
\end{figure}
\subsection{Diode Efficiencies}

We first consider a symmetric junction whose parameters are the same as in Fig.~\ref{fig:colormap_ch_sp_curr_and_spon_curr_diode_eff_SYMM} but in which both spin-channels have the same barrier width of $0.6\lambda_F/2\pi$. We then vary the temperature that leads to a variation of the superconducting order parameter $\Delta(T)$ with temperature.
Fig.~\ref{fig:appendix:symm:Diode_eff_var_temp} shows the charge [panel (a)] and the spin diode efficiency [panel (b)] as a function the geometric phase $\Delta\varphi$ and the temperature $T$. We obtain that for lower temperatures up to $T \approx 0.2 T_c$ there is a significant diode effect in both, the charge and the spin current. Moreover, $\eta_\cc$ increases reaching its maximum around $T \approx 0.2 T_c$. On the other hand, $\Delta\varphi$ at which $\eta_\sn = \pm 1$, i.e., $\widehat{\Delta\varphi}_\sn$ [see Sec.~\ref{sec:analytic_model}], shifts from $k\pi$ towards $(k+\frac{1}{2})\pi$ as $T$ approaches $0.2 T_c$. This shift can be understood by considering the approximations in section~\ref{sec:analytic_model} and Fig.~\ref{fig:B_2A_ratio_effs}(a) where we show the $B/2A$ ratio for the results shown here in Fig.~\ref{fig:appendix:symm:Diode_eff_var_temp}. From Eq.~\eqref{eq:phi_sp} follows for $|C| \ll |A|,|B|$, which is the case here, that $\widehat{\Delta\varphi}_\sn \approx \mp \arcsin\qty(B/2A)$. Thus, for low temperatures where $B < 2A$ we get $\widehat{\Delta\varphi}_\sn \approx 0$ and as $B/2A$ increases $\widehat{\Delta\varphi}_\sn \rightarrow \pi/2$. If $B > 2A$, then there exists no $\Delta\varphi$ which solves Eq.~\eqref{condition} and thus $\eta_\sn^\mathrm{max}$ decreases for $B/2A > 1$, i.e., for temperatures exceeding $T\approx 0.2 T_c$.

\begin{figure}[t!]
    \centering
    \includegraphics[width=\linewidth]{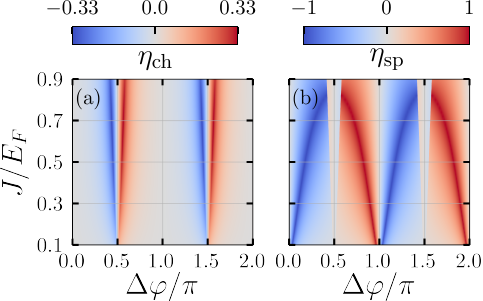}
    \caption{The functional dependence of the charge [panel (a)] and spin [panel (b)] diode efficiencies $\eta_{\cc/\sn}$ of a symmetric Josephson junction on the geometric phase $\Delta\varphi$ and the exchange field strength $J$ which is assumed to be the same in the FI layers and the sFM. In the sFM no bias potential is applied, i.e., $V_\mathrm{sFM} = 0$, whereas in the FI layer $V_\mathrm{B}^\uparrow = 1.1 E_F$ and thus $V_\mathrm{B}^\downarrow = V_\mathrm{B}^\uparrow + J$. The left and right FI layer are for both spin-channels $0.6\lambda_F /2\pi$ wide, the length of the sFM is $\xi$ and the temperature is $T=0.1T_c$.}
    \label{fig:appendix:symm:Diode_eff_var_J_VF_0}
\end{figure}

Another parameter which appears to be important for the effect is the exchange field strength in the ferromagnetic insulating layers and the metallic ferromagnet which are here, for definiteness, assumed to be equal ($J_\mathrm{B} = J_{\mathrm{sFM}} = J$). To study the effect of varying exchange field strength, in what follows we set the temperature to $T = 0.1 T_c$ and vary the exchange field within the range $0.1 < J/E_F < 0.9$. Moreover, we for now consider the case in which $V_\mathrm{sFM} = 0$, i.e., $V_\mathrm{sFM}^{\uparrow(\downarrow)} = \mp J/2$ and $k_F^\downarrow < k_F < k_F^\uparrow$, and $V_\mathrm{B}^\uparrow$ is fixed to $V_\mathrm{B}^{\uparrow} = 1.1 E_F$ and therefore $V_\mathrm{B}^\downarrow = V_\mathrm{B}^\uparrow + J$. Figure~\ref{fig:appendix:symm:Diode_eff_var_J_VF_0} shows $\eta_\cc$ [panel (a)] and $\eta_\sn$ [panel (b)] as functions of $\Delta\varphi$ and $J$. 
Apparently, the diode effect is present for all considered exchange field strengths but the functional dependencies and the magnitudes are affected by varying $J$. The functional dependence of the charge diode efficiency is unaffected by the exchange field strength but its magnitude increases with increasing exchange field strength until the maximum is reached around $J \approx 0.7 E_F$ and starts to decreases for further increment of the exchange field strength. This effect is not visible in the picture due to the colormap. In contrast, we obtain that the functional dependence of $\eta_\sn(\Delta\varphi)$ is significantly affected as the maxima and minima shift from $k\pi$ towards $(k+\frac{1}{2})\pi$. This shift can be reasoned in the same way as done for the temperature dependence and the corresponding $B/2A$ ratio and $\eta_{\cc,\sn}^\mathrm{max}$ are shown in Fig.~\ref{fig:B_2A_ratio_effs}(b). Furthermore, when $B > 2A$, i.e., $J \gtrsim 0.8 E_F$, we also obtain that $\eta_\sn^\mathrm{max} < 1$ in agreement with Eq.~\eqref{condition}.

\begin{figure}[t!]
    \centering
    \includegraphics[width=\linewidth]{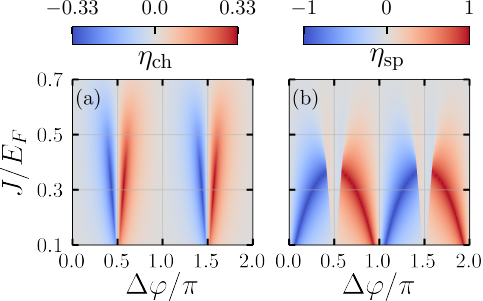}
    \caption{The functional dependence of the charge [panel (a)] and the spin [panel (b)] diode efficiencies $\eta_{\cc/\sn}$ of a symmetric Josephson junction on the geometric phase $\Delta\varphi$ and the exchange field strength $J$. Here the $\uparrow$-band in the sFM is fixed to $V_\mathrm{sFM}^\uparrow = 0.1 E_F$ and the $\downarrow$-band is shifted by $J$, i.e., $V_\mathrm{sFM}^\downarrow = V_\mathrm{sFM}^\uparrow + J$. The barrier potentials are then given by $V_\mathrm{B}^{\uparrow(\downarrow)} = E_F + V_\mathrm{sFM}^{\uparrow(\downarrow)}$. The remaining parameters are the same as for Fig.~\ref{fig:appendix:symm:Diode_eff_var_J_VF_0}.}
    \label{fig:appendix:symm:Diode_eff_var_J_simul}
\end{figure}

In addition to the case for which $V_\mathrm{sFM} = 0$ we have studied the case in which $V_\mathrm{sFM}$ is $J$-dependent such that $V_\mathrm{sFM}^\uparrow$ is fixed to $V_\mathrm{sFM}^\uparrow = 0.1 E_F$ but $V_\mathrm{sFM}^\downarrow$ varies with the exchange field $V_\mathrm{sFM}^\downarrow = V_\mathrm{sFM}^\uparrow + J$. In this case the barrier potentials are given by $V_\mathrm{B}^{\uparrow(\downarrow)} = V_\mathrm{sFM}^{\uparrow(\downarrow)} + E_F$. The functional dependencies, shown in Fig.~\ref{fig:appendix:symm:Diode_eff_var_J_simul}, differ from Fig.~\ref{fig:appendix:symm:Diode_eff_var_J_VF_0} but we obtain a similar effect as obtained for the previous case for which the same reasoning applies. Only the exact exchange fields, at which the maximum magnitude of the charge diode effect and the exchange field at which the perfect spin diode effect vanishes, differ. Moreover, the maximum diode efficiencies quickly decrease towards zero for larger exchange fields. This is because the $\downarrow$-band is then close the Fermi level and if the band becomes insulating, i.e., $J>0.9E_F$, the strongly spin-polarized ferromagnet becomes a half-metal. In the half-metallic case all contributions in which the $\downarrow$-band is involved vanish, i.e., $I_{\mu,\nu} = 0$ for $\nu \ne 0$. Therefore, in the half-metallic case $I_{\cc,\sn} = I_\upup = \sum_{\mu=1}^\infty \mu (-1)^\mu I_{\mu,0} \sin\qty[\mu\qty(\Delta\chi - \Delta\varphi)]$ and thus the diode effect is absent as positive and negative critical currents are always of the same magnitude in that case.

\begin{figure}[b]
    \centering
    \includegraphics[width=\linewidth]{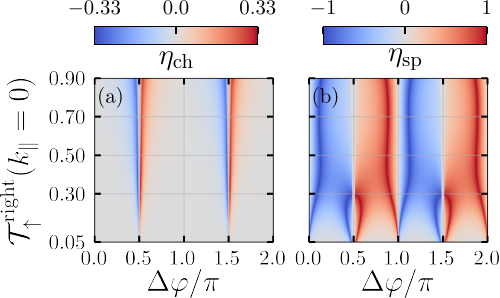}
    \caption{The functional dependence of the charge [panel (a)] and spin [panel (b)] diode efficiencies $\eta_{\cc/\sn}$ on the geometric phase difference $\Delta\varphi$ and the transmission probability of the right interface $\mathcal{T}^\mathrm{right}_\uparrow(k_\parallel=0)$, with that of the left interface $\mathcal{T}^\mathrm{left}_\uparrow(k_\parallel=0)$ fixed at 0.90. The potentials and exchange fields are as for Fig.~\ref{fig:colormap_ch_sp_curr_and_spon_curr_diode_eff_SYMM} just the spin-resolved barrier widths of the left FI layer are now both $0.60\lambda_F/2\pi$ wide.}
    \label{fig:appendix:asymm:Diode_eff_var_transm}
\end{figure}

Finally, we consider in Fig.~\ref{fig:appendix:asymm:Diode_eff_var_transm} the effect of the FI layer's width $d$ which allows to look at the crossover between a symmetric junction to an asymmetric one. To see the effect, we consider the diode efficiencies as functions of the right FI layer's widths whereas the left one is fixed at $0.60\lambda_F/2\pi$ corresponding to the transmission for perpendicular impact ($k_\parallel=0$) of $\mathcal{T}^\mathrm{left}_\uparrow \approx 0.9 $. Intuitively, it is clear that $d$ is related to the transmission probability of particles through the barrier therefore we focus here on the latter. Figure~\ref{fig:appendix:asymm:Diode_eff_var_transm} shows the charge [panel (a)] and the spin [panel (b)] diode efficiency as a function of the geometric phase difference $\Delta\varphi$ and the spin-up transmission coefficient of the right FI layer. For a more detailed analysis we consider the analytic model from Sec.~\ref{sec:analytic_model}. The maximum charge diode efficiency is reached at $B=\sqrt{2}A$ and the functional dependence is unaffected by changing the transmission of the right interface. In contrast, the functional dependence of the spin diode efficiency is strongly altered as it peaks at both $k\pi$ and $\qty(k+\frac{1}{2})\pi$ for $\mathcal{T}_\uparrow^\mathrm{right}(k_\parallel=0) \leq 0.38$ and only at $k\pi$ for a more transparent interface. The additional peaks follow from Eq.~\eqref{condition} which in the considered case has a solution close to $k\pi$ [see Eq.~\eqref{eq:phi_sp}] and another solution close to $(k+\frac{1}{2})\pi$ [see Eq.~\eqref{aroundpihalf}].

\begin{figure}[t]
    \centering
    \includegraphics[width=\linewidth]{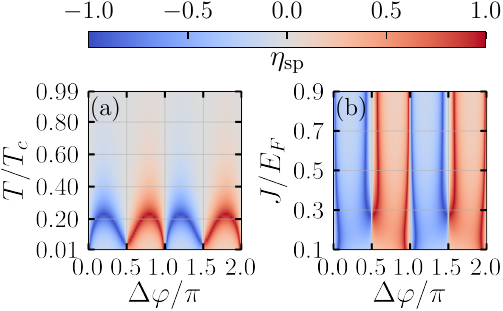}
    \caption{Functional dependence of the spin diode efficiency $\eta_\sn$ on the geometric phase $\Delta\varphi$ for varying temperature (a) and exchange field strength (b) [Fermi surface mismatch as shown in Fig.~\ref{fig:sketch_parabolas}(b)]. For both panels the barriers of the two spin-channels are of the same width for each side. The left FI layer is $0.6\lambda_F/2\pi$ wide and the right FI layer is $3.12\lambda_F/2\pi$ wide, i.e., $\mathcal{T}_\uparrow^\mathrm{left}(k_\parallel=0) \approx 0.90$ and $\mathcal{T}_\uparrow^\mathrm{right}(k_\parallel=0) \approx 0.20$. For panel (a) the exchange fields and potentials are as for Fig.~\ref{fig:appendix:symm:Diode_eff_var_temp} and for panel (b) the temperature is $T=0.1T_c$.}
    \label{fig:appendix:asymm:eta_sp_asymm_T_and_J_V_0}
\end{figure}
In Fig.~\ref{fig:appendix:asymm:eta_sp_asymm_T_and_J_V_0} we consider the spin diode efficiency in the case of an asymmetric junction where the left interface is of high transparency and the right interface is less transparent but not in the tunneling limit. We obtain that perfect efficiency is absent for $T\gtrsim 0.2 T_c$ [see Fig.~\ref{fig:appendix:asymm:eta_sp_asymm_T_and_J_V_0}(a)], but in contrast to the symmetric case [see Fig. ~\ref{fig:appendix:symm:Diode_eff_var_temp}] the geometric phase at which the maximum is reached for $T \approx 0.2 T_c$ is between successive multiples of $\pi/2$. Furthermore, it follows that the the two solutions of Eq.~\eqref{eq:phi_sp} converge to one for $T\rightarrow 0.2 T_c$. This effect can be understood in terms of Eq.~\eqref{condition}, which is a cubic polynomial in $\sin(\Delta\varphi)$. Its discriminant is zero when Eq.~\eqref{discriminant} holds. In the present case, $|A|\ll |C|$, such that this condition reads $|B|=4|C|/\sqrt{27}\approx 0.77 |C|$. The value of $\Delta\varphi$ where this condition is met is given by
Eq.~\eqref{discriminantvalue},
which for $|A|\ll |C|$ is approximately at $0.2\pi+k\pi$ and $0.8\pi+k\pi$.
Our Fourier analysis confirms that this condition is met at $T\approx 0.22 T_c$. Therefore, two solutions disappear when the temperature increases beyond this threshold. As can be seen in Fig.~\ref{fig:appendix:asymm:eta_sp_asymm_T_and_J_V_0}(a), for temperatures near this threshold, there are extended ranges in $\Delta\varphi$ for which the spin diode efficiency is close to 100\%.  Fig.~\ref{fig:appendix:asymm:eta_sp_asymm_T_and_J_V_0}(b), which is for $T=0.1T_c$, shows that by varying the exchange field a change in the number of $\Delta\varphi$ values with perfect efficiency takes place around $J\approx0.27 E_F$. At $J=0.4E_F$, corresponding to panel (a), solutions near $\Delta\varphi=(k+\frac{1}{2})\pi$ are present, whereas for $J=0.1 E_F$ these are absent. From our discussion of Eq.~\eqref{aroundpihalf} the condition $|B|>2|A|$ must be met for perfect efficiency around $\Delta\varphi=(k+\frac{1}{2})\pi$, and as the ratio $|B|/2|A|$ increases with $J/E_F$ (it approaches zero for $J\to 0$), there is a critical $J/E_F \approx 0.27$ which must be exceeded for these solutions to exist. In Fig.~\ref{fig:appendix:asymm:eta_sp_asymm_T_and_J_V_0}(b) this is clearly visible.

\subsection{Spontaneous Currents}
Beside the diode efficiencies an interesting effect present in a SC/sFM/SC trilayer is the spontaneous current which is induced by the geometric phase difference $\Delta\varphi$. 
We concentrate on the spontaneous currents in asymmetric junctions because in this case the spontaneous spin current can be dominated by a second harmonic contribution. In our calculations,
the barrier widths are spin-degenerate but mutually different: the left FI layer has high transparency, $d_L = 0.6 \lambda_F / 2\pi$, and the right FI layer is much less transparent, $d_R = 4.16 \lambda_F / 2\pi$. 

\begin{figure}[b]
    \centering
    \includegraphics[width=\linewidth]{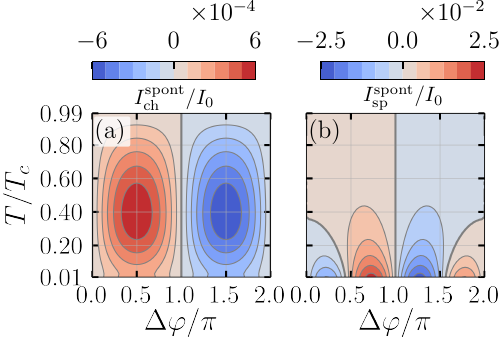}
    \caption{Spontaneous charge [panel (a)] and spin [panel (b)] currents $I_{\cc/\sn}^\mathrm{spont}$ as functions of geometric phase difference $\Delta\varphi$ and temperature $T$ for an asymmetric junction with barrier potentials of height $V_\mathrm{B}^\uparrow = 1.1 E_F$ and  $V_\mathrm{B}^\downarrow = 1.7 E_F$ and the bands in the sFM are at $V_\mathrm{sFM}^{\uparrow(\downarrow)} = \mp 0.3 E_F$. The FI layer widths are spin-degenerate with $d_L = 0.6\lambda_F/2\pi$ at the left interface and $d_R = 4.16\lambda_F / 2\pi$ at the right interface.}
    \label{fig:appendix:asymm:Spon_curr_var_temp}
\end{figure}

In Fig.~\ref{fig:appendix:asymm:Spon_curr_var_temp} we present the spontaneous charge [panel (a)] and spin [panel (b)] currents as functions of the temperature $T$ and the geometric phase difference $\Delta\varphi$. 
The spontaneous charge current is well-approximated by $I_\cc(\Delta\chi = 0,\Delta\varphi,T) \approx I_\cc^{(0)}(T) \sin(\Delta\varphi)$ where $I_\cc^{(0)}(T)$ is maximum for $T\approx 0.4 T_c$ and decreases for smaller or higher temperatures. In contrast, the spontaneous spin current shows a strong second harmonic at low temperatures which decreases with increasing temperatures. With $T$ approaching $T_c$, it becomes purely sinusoidal.

\begin{figure}[t]
    \centering
    \includegraphics[width=\linewidth]{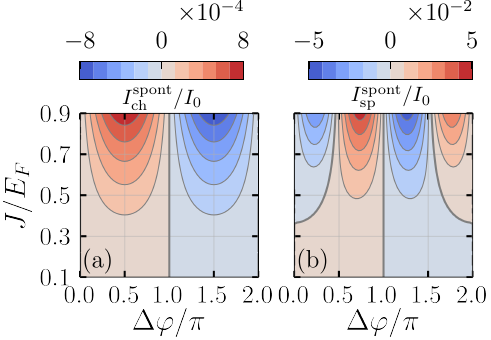}
    \caption{Spontaneous charge [panel (a)] and spin [panel (b)] currents $I_{\cc/\sn}^\mathrm{spont}$ as a function of the geometric phase difference $\Delta\varphi$ and the exchange field strength $J$ which is assumed to be the same in the FI layer and the sFM, i.e. as in Fig.~\ref{fig:appendix:symm:Diode_eff_var_J_VF_0}. The FI layer widths are as for Fig.~\ref{fig:appendix:asymm:Spon_curr_var_temp}, i.e., for an asymmetric junction, and the temperature is $T=0.1T_c$.}
    \label{fig:appendix:asymm:Spon_curr_var_J}
\end{figure}

Fig.~\ref{fig:appendix:asymm:Spon_curr_var_J} shows the spontaneous charge [panel (a)] and spin [panel (b)] currents as functions of the exchange field $J$ and the geometric phase difference $\Delta\varphi$ for fixed temperature to $T=0.1T_c$. The exchange field is assumed to have the same magnitude in the ferromagnetic insulating layers and in the metallic ferromagnet.
Similarly to the case discussed above, the spontaneous charge current is approximately sinusoidal as function of $\Delta\varphi$, with the pre-factor $I_\cc^{(0)}(J)$ a monotonically increasing function of $J$. In contrast, the spontaneous spin current is strongly affected by $J$. 
For small $J$ it is approximately sinusoidal in $\Delta\varphi$, whereas a second harmonic appears and becomes pronounced as the exchange field increases.

\section{Density of states}
\begin{figure}[b]
    \centering
    \includegraphics[width=0.9\linewidth]{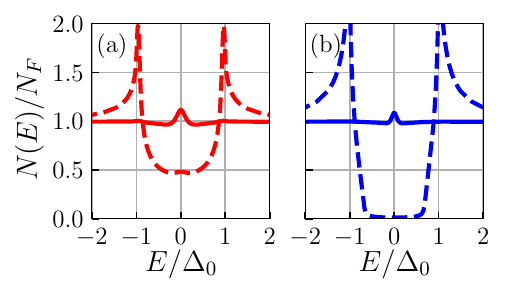}
    \caption{Local density of states (LDOS) $N(E)$ (a) of the symmetric junction and (b) of the asymmetric junction. The solid lines show the LDOS in the center of the sFM, i.e., $z = 5.5\xi$ and the dashed lines the LDOS at the superconducting side of the right interface, i.e., $z = 6\xi$. The superconducting and geometric phase difference are set to zero. The parameters of the symmetric and asymmetric case are the same as for Figs.~\ref{fig:colormap_ch_sp_curr_and_spon_curr_diode_eff_SYMM} and \ref{fig:colormap_sp_curr_diode_eff_ASYMM}, respectively.}
    \label{fig:appendix:DOS_comparison}
\end{figure}
The knowledge of the quasiclassical Green's function is sufficient to express most of relevant physical quantities. In particular, it allows calculating the spectral properties such as the local density of states (LDOS) besides the transport properties. Having obtained the quasiclassical Green's function, the LDOS per spin projection, $N(z,E)$, is calculated via \cite{belzigQuasiclassicalGreensFunction1999}
\begin{align}
    N_\eta (z,E) &= N_{F\eta}\, \Re\left[\mathcal{G}_{\eta\eta}(z,E)\right], \\
    N(z,E)&=\frac{1}{2}\left[N_\uparrow(z,E) +N_\downarrow(z,E)\right],
\end{align}
where $N_{F\eta}$ is the spin-resolved density of states at the Fermi level in the normal state.

In Fig.~\ref{fig:appendix:DOS_comparison} we show the self-consistently obtained LDOS as a function of
 energy for (a) the symmetric junction considered in Fig.~\ref{fig:colormap_ch_sp_curr_and_spon_curr_diode_eff_SYMM} and (b) the asymmetric junction considered in Fig.~\ref{fig:colormap_sp_curr_diode_eff_ASYMM}. In both panels, the solid lines correspond to the LDOS in the middle of the sFM and the dashed ones to the LDOS on the superconducting side of the right interface, i.e., sFM/SC interface. The zero-energy peaks visible in the solid lines in both panels is a result of the spin-triplet nature of the pair amplitudes in the sFM, which in the diffusive limit are all three odd-frequency correlations~\cite{bergeretOddTripletSuperconductivity2005,yokoyamaResonantPeakDensity2005,yokoyamaManifestationOddfrequencySpintriplet2007,tanakaTheoryProximityEffect2007,dibernardoSignatureMagneticdependentGapless2015,linderOddfrequencySuperconductivity2019}. Instead of a minigap, expected in a conventional Josephson junction with a normal metal between two superconductors, here an enhanced density of states is observed near zero bias. 
 The LDOS in the superconductor displays the coherence peaks close to the superconducting gap but still being far from the BCS density of states: the gap is either partly filled in the symmetric junction [panel (a)] or reduced in the asymmetric one [panel (b)]. For both junction types we recover the BCS density of states deep in the superconductors, i.e., close to the outer interfaces.

\end{document}